\documentclass[a4paper,aps,pre,floatfix,twocolumn,nofootinbib,showpacs,superscriptaddress]{revtex4-1}

\usepackage[pdftex]{graphicx}
\setkeys{Gin}{width=0.9\columnwidth}
\usepackage{latexsym,amsmath}
\usepackage[pdftex]{color}
\usepackage{hyperref}
\usepackage{float}
\usepackage[all]{xy}
\usepackage{verbatim}
\usepackage{lipsum}

\newcommand{\av}[1]{\left\langle #1 \right\rangle}

\graphicspath{{../figures/}}

\begin{document}

\title{Effect of risk perception on epidemic spreading in temporal
  networks}

\date{\today}
\author{Antoine Moinet}

\affiliation{Aix Marseille Univ, Universit\'e de Toulon, CNRS, CPT,
  Marseille, France}

\affiliation{Departament de Fisica, Universitat Politecnica de
  Catalunya, Campus Nord B4, 08034 Barcelona, Spain}

\author{Romualdo Pastor-Satorras}

\affiliation{Departament de Fisica, Universitat Politecnica de
  Catalunya, Campus Nord B4, 08034 Barcelona, Spain}

\author{Alain Barrat}

\affiliation{Aix Marseille Univ, Universit\'e de Toulon, CNRS, CPT,
  Marseille, France}

\affiliation{Data Science Laboratory, ISI Foundation, Torino, Italy}

\begin{abstract}
  Many progresses in the understanding of epidemic spreading models have
  been obtained thanks to numerous modeling efforts and analytical and
  numerical studies, considering host populations with very different
  structures and properties, including complex and temporal interaction
  networks. Moreover, a number of recent studies have started to go
  beyond the assumption of an absence of coupling between the spread of
  a disease and the structure of the contacts on which it unfolds.
  Models including awareness of the spread have been proposed, to mimic
  possible precautionary measures taken by individuals that decrease
  their risk of infection, but have mostly considered static
  networks. Here, we adapt such a framework to the more realistic case
  of temporal networks of interactions between individuals. We study the
  resulting model by analytical and numerical means on both simple
  models of temporal networks and empirical time-resolved contact
  data. Analytical results show that the epidemic threshold is not
  affected by the awareness but that the prevalence can be significantly
  decreased. Numerical studies highlight however the presence of very
  strong finite-size effects, in particular for the more realistic
  synthetic temporal networks, resulting in a significant shift of the
  effective epidemic threshold in the presence of risk awareness. For
  empirical contact networks, the awareness mechanism leads as well to a
  shift in the effective threshold and to a strong reduction of the
  epidemic prevalence.
\end{abstract}

\maketitle

\section{Introduction}

The propagation patterns of an infectious disease depend on many
factors, including the number and properties of the different stages of
the disease, the transmission and recovery mechanisms and rates, and the
hosts' behavior (e.g., their contacts and
mobility)~\cite{Keeling-Rohani_Book,Anderson-May_Book}. Given the
inherent complexity of a microscopic description taking into account all
details, simple models are typically used as basic mathematical
frameworks aiming at capturing the main characteristics of the epidemic
spreading process and in particular at understanding if and how
strategies such as quarantine or immunization can help contain it. Such
models have been developed with increasing levels of sophistication and
detail in the description of both the disease evolution and the
behaviour of the host
population~\cite{Keeling-Rohani_Book,Anderson-May_Book}.

The most widely used assumption concerning the disease evolution within
each host consists in discretizing the possible health status of
individuals~\cite{Keeling-Rohani_Book,Anderson-May_Book}. For instance,
in the Susceptible-Infectious-Susceptible (SIS) model, each individual
is considered either healthy and susceptible (S) or infectious (I).
Susceptible individuals can become infectious through contact with an
infectious individual, and recover spontaneously afterwards, becoming
susceptible again.  In the Susceptible-Infectious-Recovered (SIR) case,
recovered individuals are considered as immunized and cannot become
infectious again.  The rate of infection during a contact is assumed to
be the same for all individuals, as well as the rate of recovery.

Obviously, the diffusion of the disease in the host population depends
crucially on the patterns of contacts between hosts.  The simplest
homogeneous mixing assumption, which makes many analytical results
achievable, considers that individuals are identical and that each has a
uniform probability of being in contact with any other
individual~\cite{Keeling-Rohani_Book,Anderson-May_Book}.  Even within
this crude approximation, it is possible to highlight fundamental
aspects of epidemic spreading, such as the epidemic threshold, signaling
a non-equilibrium phase transition that separates an epidemic-free phase
from a phase in which a finite fraction of the population is affected
\cite{Keeling-Rohani_Book}.  However, this approach neglects any
non-trivial structure of the contacts effectively occurring within a
population, while advances in network science~\cite{Newman10} have shown
that a large number of networks of interest have in common important
features such as a strong heterogeneity in the number of connections, a
large number of triads, a community structure, and a low average
shortest path length between two individuals
\cite{caldarelli2007sfn,Newman10}.  Spreading models have thus been
adapted to complex networks, and studies have unveiled the important
role of each of these properties
\cite{Pastor-Satorras:2001,BBV,Pastor-Satorras:2015}.  More recently, a
number of studies have also considered spreading processes on
time-varying networks
\cite{Holme:2012,Karsai:2011,Stehle:2011,Machens:2013,Valdano:2015,Holme:2015},
to take into account the fact that contact networks evolve on various
timescales and present non-trivial temporal properties such as broad
distribution of contact durations \cite{Cattuto:2010,Barrat:2014} and
burstiness \cite{Barabasi:2005,Holme:2012} (i.e., the timeline of social
interactions of a given individual exhibits periods of time with intense
activity separated by long quiescent periods with no interactions).

All these modeling approaches consider that the propagation of the
disease takes place on a substrate (the contacts between individuals)
that does not depend on the disease itself. In this framework, standard
containment measures consist in the immunization of individuals, in
order to effectively remove them from the population and thus break
propagation paths. Immunization can also (in models) be performed in a
targeted way, trying to identify the most important (class of) spreaders
and to suppress propagation in the most efficient possible
way~\cite{Pastor:2002,Cohen:2003}.  An important point to consider
however is that the structure and properties of contacts themselves can
in fact be affected by the presence of the disease in the population, as
individuals aware of the disease can modify their behaviour in
spontaneous reaction in order to adopt self-protecting measures such as
vaccination or mask-wearing.  A number of studies have considered this
issue along several directions (see Ref.~\cite{Funk:2010} for a
review). For instance, some works consider an adaptive evolution of the
network~\cite{Gross:2008} with probabilistic redirection of links
between susceptible and infectious individuals, to mimic the fact that a
susceptible individual might be aware of the infectious state of some of
his/her neighbors, and therefore try to avoid contact with them.

Other works introduce behavioral classes in the population, depending on
the awareness to the disease \cite{Perra:2011}, possibly consider that
the awareness of the disease propagates on a different (static) network
than the disease itself, and that being aware of the disease implies a
certain level of immunity to it
\cite{Granell:2013,Massaro:2014}. Finally, the fact that an individual
takes self-protecting measures that decrease his/her probability to be
infected (such as wearing a mask or washing hands more frequently) can
depend on the fraction of infectious individuals present in the whole
population or among the neighbors of an individual. These measures are
then modeled by the fact that the probability of a susceptible catching
the disease from an infectious neighbor depends on such fractions
\cite{Bagnoli:2007,Kotnis:2013,Rizzo:2014,Cao:2014}.  Yet these studies
mostly consider contacts occurring on a static underlying contact
network (see however \cite{Kotnis:2013,Rizzo:2014} for the case of a
temporal network in which awareness has the very strong effect of
reducing the activity of individuals and their number of contacts,
either because they are infectious or because of a global knowledge of
the overall incidence of the disease).

Here, we consider instead the following scenario: First, individuals are
connected by a time-varying network of contacts, which is more realistic
than a static one; second, we use the scenario of a relatively mild
disease, which does not disrupt the patterns of contacts but which leads
susceptible individuals who witness the disease in other individuals to
take precautionary measures. We do not assume any knowledge of the
overall incidence, which is usually very difficult to know in a real
epidemic, especially in real time.  We consider SIS and SIR models and
both empirical and synthetic temporal networks of contacts. We extend
the concept of awareness with respect to the state of neighbors from
static to temporal networks and perform extensive numerical simulations
to uncover the change in the phase diagram (epidemic threshold and
fraction of individuals affected by the disease) as the parameters
describing the reaction of the individuals are varied.

\section{Temporal networks}

We will consider as substrate for epidemic propagation both synthetic
and empirical temporal networks of interactions. We describe them
succinctly in the following Subsections.

\subsection{Synthetic networks}
\label{sec:synthetic-networks}

\subsubsection{Activity-driven network model}

The activity driven (AD) temporal network model proposed in
Ref.~\cite{Perra:2012} considers a population of $N$ individuals
(agents), each agent $i$ characterized by an activity potential $a_i$,
defined as the probability that he/she engages in a social
act/connection with other agents per unit time. The activity of the
agents is a (quenched) random variable, extracted from the activity
potential distribution $F(a)$, which can take a priori any form.  The
temporal network is built as follows: at each time step $t$, we start
with $N$ disconnected individuals. Each individual $i$ becomes active
with probability $a_i$. Each active agent generates $m$ links (starts
$m$ social interactions) that are connected to $m$ other agents selected
uniformly at random (among all agents, not only active ones)
\footnote{Note that with such a definition, an agent may both receive
  and emit a link to the same other agent. However, we consider here an
  unweighted and undirected graph, thus in such a case, a single link is
  considered.  Moreover, in the limit of large $N$, the probability of
  such an event goes to $0$.}.  The resulting set of $N$ individuals and
links defines the instantaneous network $G_t$. At the next time step,
all links are deleted and the procedure is iterated. For simplicity, we
will here consider $m=1$.

In Ref.~\cite{Perra:2012} it was shown that several empirical networks
display broad distributions of node activities, with functional shapes
close to power-laws for $F(a)$, with exponents between $2$ and $3$. The
aggregation of the activity-driven temporal network over a time-window
of length $T$ yields moreover a static network with a long-tailed degree
distribution of the form $P_T(k) \sim F(k/T)$
\cite{Perra:2012,PhysRevE.87.062807}.  Indeed, the individuals with the
highest activity potential tend to form a lot more connections than the
others and behave as hubs, which are known to play a crucial role in
spreading processes~\cite{Pastor-Satorras:2015}.

\subsubsection{Activity-driven network model with memory}

A major shortcoming of the activity-driven model lies in the total
absence of correlations between the connections built in successive time
steps.  It is therefore unable to reproduce a number of features
observed in empirical data. An extension of the model tackles this issue
by introducing a memory effect into the mechanism of link
creation~\cite{Karsai:2014}. In the resulting activity-driven model with
memory (ADM), each individual keeps track of the set of other
individuals with whom there has been an interaction in the past. At each
time step $t$ we start as in the AD model with $N$ disconnected
individuals, and each individual $i$ becomes active with probability
$a_i$. For each link created by an active individual $i$, the link goes
with probability $p=q_i(t)/[q_i(t) + 1]$ to one of the $q_i(t)$
individuals previously encountered by $i$, and with probability $1-p$
towards a never encountered one. In this way, contacts with already
encountered other individuals have a larger probability to be repeated
and are reinforced. As a result, for a power-law distributed activity
$F(a)$, the degree distribution of the temporal network aggregated on a
time window $T$ becomes narrow, while the distribution of weights
(defined as the number of interactions between two individuals) becomes
broad \cite{Karsai:2014}.

\subsection{Empirical social networks}
\label{sec:empir-soci-netw}

In addition to the simple models described above, which do not exhibit
all the complexity of empirical data, we also consider two datasets
gathered by the SocioPatterns collaboration \cite{SocioPatterns}, which
describe close face-to-face contacts between individuals with a temporal
resolution of $20$ seconds in specific contexts (for further details,
see Ref.~\cite{Cattuto:2010}).  We consider first a dataset describing
the contacts between students of nine classes of a high school (Lyc\'ee
Thiers, Marseilles, France), collected during $5$ days in Dec. 2012
(``Thiers'' dataset)~\cite{SchoolDataset,Fournet:2014}. We also use
another dataset consisting in the temporal network of contacts between
the participants of a conference (2009 Annual French Conference on
Nosocomial Infections, Nice, France) during one day (``SFHH'' dataset)
\cite{Stehle:2011}. The SFHH (conference) data correspond to a rather
homogeneous contact network, while the Thiers (high school) population
is structured in classes of similar sizes and presents contact patterns
that are constrained by strict and repetitive school schedules. In
Table~\ref{tab:summary} we provide a brief summary of the main
properties of these two datasets.

\begin{table}[b]
  \begin{ruledtabular} 
    \begin{tabular}{|c||c|c||c|c||c|c|}
    Dataset & $N$ & $T$  & $\overline{p}$ &$\av{\Delta t}$ & $\av{k}$ & $\av{s}$ \\ \hline  
      Thiers  &  180   & 14026   & 5.67  & 2.28 & 24.66 & 500.5 \\ 
      SFHH     &  403   & 3801   & 26.14  & 2.69 & 47.47 & 348.7\\ 
    \end{tabular}
  \end{ruledtabular} 
  \caption{Some properties of the SocioPatterns datasets under
    consideration: $N$, number of different individuals engaged in
    interactions; $T$, total duration of the contact sequence, in units
    of the elementary time interval $t_0 = 20$ seconds; $\overline{p}$,
    average number of individuals interacting at each time step;
    $\av{\Delta t}$, average duration of a contact; $\av{k}$ and
    $\av{s}$: average degree and average strength of the nodes in the
    network aggregated over the whole time sequence.}
  \label{tab:summary}
\end{table}

\section{Modelling epidemic spread in temporal networks}
\subsection{Epidemic models and epidemic threshold}
\label{sec:epid-models-epid}

We consider the paradigmatic Susceptible-Infectious-Susceptible (SIS)
and Susceptible-Infectious-Recovered (SIR) models to describe the spread
of a disease in a fixed population of $N$ individuals. In the SIS model,
each individual belongs to one of the following compartments: healthy
and susceptible (S) or diseased and infectious (I). A susceptible
individual in contact with an infectious becomes infectious at a given
constant rate, while each infectious recovers from infection at another
constant rate. In the SIR case, infectious individuals enter the
recovered (R) compartment and cannot become infectious anymore. We
consider a discrete time modeling approach, in which the contacts
between individuals are given by a temporal network encoded in a
time-dependent adjacency matrix $A_{ij}(t)$ taking value $1$ if
individuals $i$ and $j$ are in contact at time $t$, and $0$
otherwise. At each time step, the probability that a susceptible
individual $i$ becomes infectious is thus given by
$p_i=1-\prod_j [1-\lambda \, A_{ij}(t)\, \sigma_j]$, where $\lambda$ is the
infection probability, and $\sigma_j$ is the state of node $j$
($\sigma_j=1$ if node $j$ is infectious and $0$ otherwise). We define
$\mu$ as the probability that an infectious individual recovers during a
time step. The competition between the transmission and recovery
mechanisms determines the epidemic threshold.  Indeed, if $\lambda$ is
not large enough to compensate the recovery process ($\lambda/\mu$
smaller than a critical value), the epidemic outbreak will not affect a
finite portion of the population, dying out rapidly. On the other hand,
if $\lambda/\mu$ is large enough, the spread can lead in the SIS model
to a non-equilibrium stationary state in which a finite fraction of the
population is in the infectious state. For the SIR model, on the other
hand, the epidemic threshold is determined by the fact that the fraction
$r_\infty=R_\infty/N$ of individuals in the recovered state at the end
of the spread becomes finite for $\lambda/\mu$ larger than the
threshold.

In order to numerically determine the epidemic threshold of the SIS
model, we adapt the method proposed in Refs.~\cite{Boguna:2013,Mata15},
which consists in measuring the lifetime and the coverage of
realizations of spreading events, where the coverage is defined as the
fraction of distinct nodes ever infected during the realization.  Below
the epidemic threshold, realizations have a finite lifetime and the
coverage goes to $0$ in the thermodynamic limit. Above threshold, the
system in the thermodynamic limit has a finite probability to reach an
endemic stationary state, with infinite lifetime and coverage going to
$1$, while realizations that do not reach the stationary state have a
finite lifetime.  The threshold is therefore found as the value of
$\lambda/\mu$ where the average lifetime of non-endemic realizations
diverges.  For finite systems, one can operationally define an arbitrary
maximum coverage $C>0$ (for instance $C=0.5$) above which a realization
is considered endemic, and look for the peak in the average lifetime of
non-endemic realizations as a function of $\lambda/\mu$.

In the SIR model the lifetime of any realization is finite. We thus
evaluate the threshold as the location of the peak of the relative
variance of the fraction $r_\infty$ of recovered individuals at the end
of the process~\cite{Castellano2016}, i.e.,
\begin{equation}
  \sigma_r = \dfrac{\sqrt{\av{r^{2}_{\infty}}-\av{r_\infty}^2}}{r_\infty}.
  \label{sigma}
\end{equation}

\subsection{Modeling risk perception}

To model risk perception, we consider the approach proposed in
Ref.~\cite{Bagnoli:2007} for static interaction networks. In this
framework, each individual $i$ is assumed to be aware of the fraction of
his/her neighbors who are infectious at each time step. This awareness
leads the individual to take precautionary measures that decrease its
probability to become infectious upon contact. This decrease is modeled
by a reduction of the transmission probability by an exponential factor:
at each time step, the probability of a susceptible node $i$ in contact
with an infectious to become infectious depends on the neighborhood of
$i$ and is given by $\lambda_i(t) = \lambda_0 \exp(-J n_i(t)/k_i)$ where
$k_i$ is the number of neighbors of $i$, $n_i(t)$ the number of these
neighbors that are in the infectious state at time $t$, and $J$ is a
parameter tuning the degree of awareness or amount of precautionary
measures taken by individuals.

Static networks of interactions are however only a first approximation
and real networks of contacts between individuals evolve on multiple
timescales \cite{Barrat:2014}. We therefore consider in the present
work, more realistically, that the set of neighbors of each individual
$i$ changes over time. We need thus to extend the previous concept of
neighborhood awareness to take into account the history of the contacts
of each individual and his/her previous encounters with infectious
individuals.  We consider that longer contacts with infectious
individuals should have a stronger influence on a susceptible
individual's awareness, and that the overall effect on any individual
depends on the ratio of the time spent in contact with infectious to the
total time spent in contact with other individuals. Indeed, two
individuals spending a given amount of time in contact with infectious
individuals may react differently depending on whether these contacts
represent a large fraction of their total number of contacts or not. We moreover argue that the awareness is
influenced only by recent contacts, as having encountered ill
individuals in a distant past is less susceptible to lead to a change of
behaviour.  To model this point in a simple way, we consider that each
individual has a finite memory of length $\Delta T$ and that only
contacts taking place in the time window $[t-\Delta T,t[$, in which the
present time $t$ is excluded, are relevant.

We thus propose the following risk awareness change of behaviour: The
probability for a susceptible individual $i$, in contact at time $t$ with an
infectious one, to become infectious, is given by
\begin{equation}
  \lambda_i(t) = \lambda_0 \exp\left(-\alpha \, n_I(i)_{\Delta T} \right) 
\label{risk}
\end{equation}
where $n_I(i)_{\Delta T}$ is the number of contacts with infectious
individuals seen by the susceptible during the interval
$[t-\Delta T,t[$, divided by the total number of contacts counted by the
individual during the same time window (repeated contacts between the
same individuals are also counted). $\alpha$ is a parameter gauging the
strength of the awareness, and the case $\alpha = 0$ corresponds to the
pure SIS process, in which $\lambda_i(t) = \lambda_0$ for all
individuals and at all times.

\section{Epidemic spreading on synthetic networks}

\subsection{SIS dynamics}
\label{sec:sis-dynamics}

\subsubsection{Analytical approach}

On a synthetic temporal network, an infectious individual can propagate
the disease only when he/she is in contact with a susceptible. As a
result, the spreading results from an interplay between the recovery
time scale $1/\mu$, the propagation probability $\lambda$ conditioned on
the existence of a contact and the multiple time scales of the network
as emerging from the distribution of nodes' activity $F(a)$. Analogously
to what is done for heterogeneous static networks
\cite{BBV,Pastor-Satorras:2015}, it is possible to describe the spread
at a mean-field level by grouping nodes in activity classes: all nodes
with the same activity $a$ are in this approximation considered
equivalent~\cite{Perra:2012}. The resulting equation for the evolution
of the number of infectious nodes in the class of nodes with activity
$a$ in the original AD model has been derived in Ref.~\cite{Perra:2012}
and reads
\begin{equation}
I_{a}^{t+1}=I_{a}^{t}- \mu \, I_{a}^{t}+ \lambda\, a\, S_{a}^{t}\int
\dfrac{I_{a'}^{t}}{N}\,da' + \lambda\, S_{a}^{t}\int
\dfrac{I_{a'}^{t}\,a'}{N}\,da'  \ . 
\label{bilan}
\end{equation}
where $I_a$ and $S_a$ are the number of infectious and susceptible nodes
with activity $a$, verifying $N_a=S_a+I_a$. 

From this equation one can show, by means of a linear stability
analysis, that there is an endemic non-zero steady state if and only if
$(\av{a}+\sqrt{\av{a^2}})\lambda/\mu > 1 $~\cite{Perra:2012}.  Noticing
that $\av{a}+\sqrt{\av{a^2}}$ may be regarded as the highest
statistically significant activity rate, the interpretation of this
equation becomes clear: the epidemic can propagate to the whole network
when the smallest time scale of relevance for the infection process is
smaller than the time scale of recovery.

Let us now consider the introduction of risk awareness in the SIS
dynamics on AD networks. In general, we can write for a susceptible with
activity $a$
\begin{equation}
  n_I(a)_{\Delta T} = \dfrac{\sum\limits_{i=1}^{\Delta T}\left(a \int
      \dfrac{I_{a'}^{t-i}}{N}\,da' 
      + \int \dfrac{I_{a'}^{t-i}\,a'}{N}\,da'\right)
  }{(a+\av{a})\,\Delta T} \ ,
  \label{eq:4}
\end{equation}
where the denominator accounts for the average number of contacts of an
individual with activity $a$ in $\Delta T$ time steps. In the steady state, where the
quantities $I_a$ become independent of $t$, the dependence on $\Delta T$
in Eq.~(\ref{eq:4}) vanishes, since both the average time in contact
with infectious individuals and the average total time in contact are
proportional to the time window width. Introducing this expression into
Eq.~(\ref{risk}), we obtain
\begin{equation}
  \label{eq:3}
  \lambda_a = \lambda_0 \exp\left( -\alpha\, \dfrac{a \int \dfrac{I_{a'}}{N}\,da'
      + \int \dfrac{I_{a'}\,a'}{N}\,da' }{a+\av{a}}  \right),
\end{equation}
which can be inserted into Eq.~(\ref{bilan}). Setting $\mu=1$ without
loss of generality, we obtain the steady state solution
\begin{equation}
  \rho_a = \dfrac{\lambda_a (a\rho+\theta)}{1+\lambda_a (a\rho+\theta)} ,
  \label{eqrhoa}
\end{equation}
where  $\rho_a = I_a / N_a$ and we have defined
\begin{eqnarray}
\rho &=& \sum_a F(a) \rho_a, \label{eq:5} \\
\theta &=& \sum_a a\,F(a)\rho_a.  \label{eq:6}
\end{eqnarray}
Introducing Eqs.~(\ref{eq:3}) and~(\ref{eqrhoa}) into Eqs.~(\ref{eq:5})
and~(\ref{eq:6}), and expanding at second order in $\rho$ and $\theta$,
we obtain after some computations the epidemic threshold
\begin{equation}
  \lambda_c = \dfrac{1}{\av{a}+\sqrt{\av{a^2}}}   .
\end{equation}
Moreover, setting $\lambda_0=\lambda_c(1+\epsilon)$ and expanding at order 1 in $\epsilon$ we obtain
\begin{equation}
  \rho = \dfrac{2\epsilon}{A\alpha+B},
\end{equation}
where
\begin{eqnarray}
  A &=& \lambda_c\av{\dfrac{\dfrac{a^3}{\sqrt{\av{a^2}}}+3a
        \sqrt{\av{a^2}}+\av{a^2}+3a^2}{a+\av{a}}}  \\ 
B &=& \lambda_{c}^{2}\left(\dfrac{\av{a^3}}{\sqrt{\av{a^2}}}+3\av{a}
      \sqrt{\av{a^2}}+4\av{a^2}\right)\nonumber . 
\end{eqnarray}
This indicates that, at the mean-field level, the epidemic threshold is
not affected by the awareness.  Nevertheless, the density of infectious
individuals in the vicinity of the threshold is reduced as the awareness
strength $\alpha$ grows. 

In the case of activity driven networks with memory (ADM), no analytical
approach is available for the SIS dynamics, even in the absence of
awareness. The numerical investigation carried out in
Ref.~\cite{Sun:2015} has shown that the memory mechanism, which leads to
the repetition of some contacts, reinforcing some links and yielding a
broad distribution of weights, has a strong effect in the SIS
model. Indeed, the repeating links help the reinfection of nodes that
have already spread the disease and make the system more
vulnerable to epidemics. As a result, the epidemic threshold is reduced with
respect to the memory-less (AD) case. For the SIS dynamics with
awareness on ADM networks, we will now resort to numerical simulations.

\subsubsection{Numerical simulations}

In order to inspect in details the effect of risk awareness on the SIS
epidemic process, we perform extensive numerical simulations.  Following
Refs.~\cite{Perra:2012,Sun:2015}, we consider a distribution of nodes'
activities of the form $F(a) \propto a^{-\gamma}$ for
$a \in [\epsilon,1]$, where $\epsilon$ is a lower activity cut-off
introduced to avoid divergences at small activity values. In all
simulations we set $\epsilon = 10^{-3}$ and $\gamma = 2$. We consider
networks up to a size $N=10^5$ and a SIS process starting with a
fraction $I_0/N=0.01$ of infectious nodes chosen at random in the
population. In order to take into account the connectivity of the
instantaneous networks, we use as a control parameter the quantity
$\beta / \mu$, where $\beta = 2 \av{a} \lambda_0$ is the per capita rate
of infection~\cite{Perra:2012}. Notice that the average degree of an
instantaneous network is $\av{k}_ t = 2 \av{a}$
\cite{PhysRevE.87.062807}. With this definition, the critical endemic
phase corresponds to
\begin{equation}
  \label{eq:1}
  \frac{\beta}{\mu} \geq \frac{2 \av{a}}{\av{a} + \sqrt{\av{a^2}}}.
\end{equation}

\begin{figure}[t]
  \centering \includegraphics{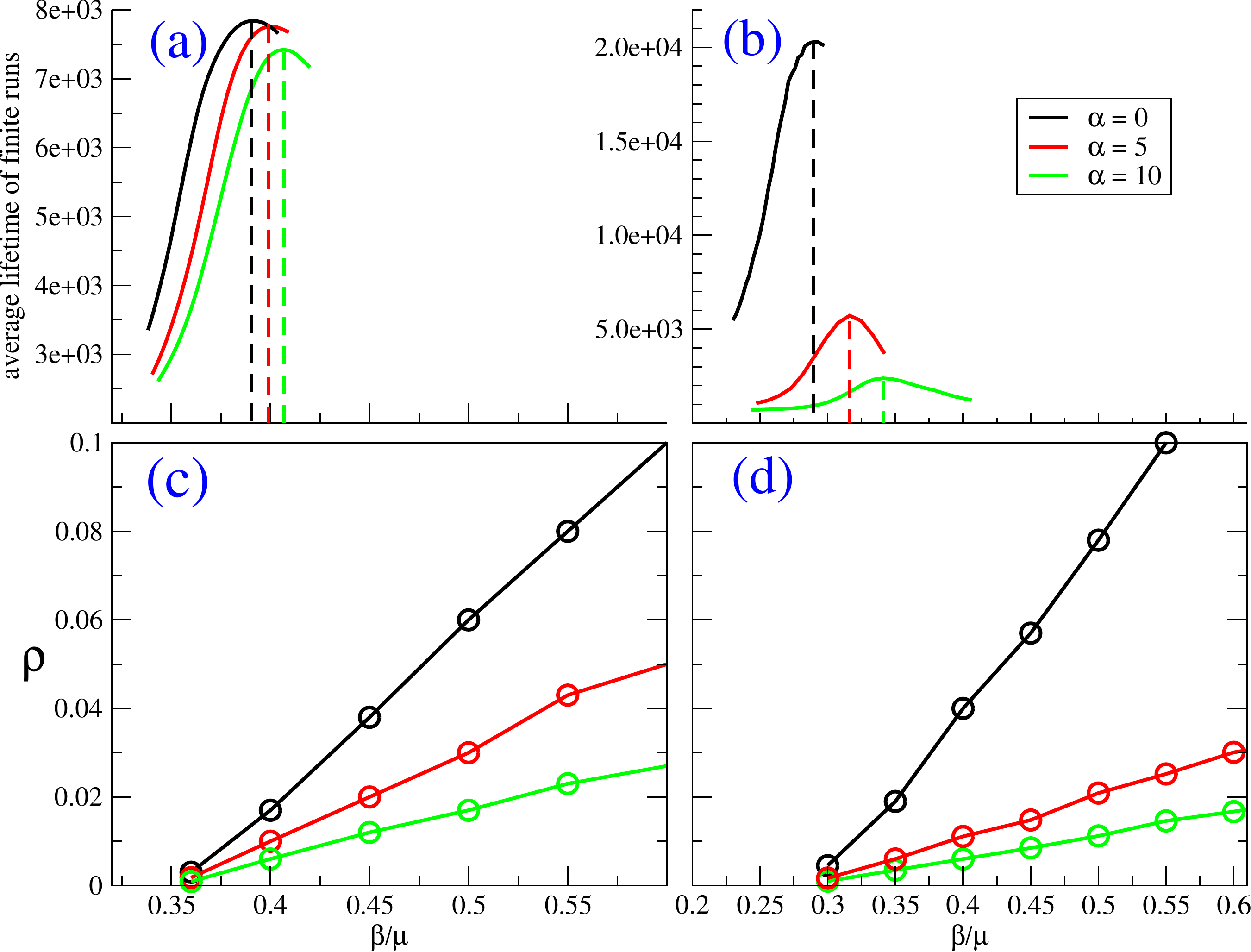}
  \caption{Effect of the strength of risk awareness on the SIS spreading
    on AD and ADM networks with $\Delta T = \infty$. (a): average
    lifetime of non-endemic runs for AD network, (b): average lifetime
    of non-endemic runs for ADM networks, (c): Steady state fraction of
    infectious for AD, (d): Steady state fraction of infectious for
    ADM. Vertical lines in subplots (a) and (b) indicate the position of
    the maximum of the average lifetime. Model parameters:
    $\mu = 0.015$, $\gamma = 2$, $\epsilon = 10^{-3}$,
    $\Delta T = \infty$ and network size $N = 10^5$. Results are
    averaged over $1000$ realizations. }
  \label{fig:ADSISM1A1}
\end{figure}

\begin{figure}[t]
  \centering
  \includegraphics{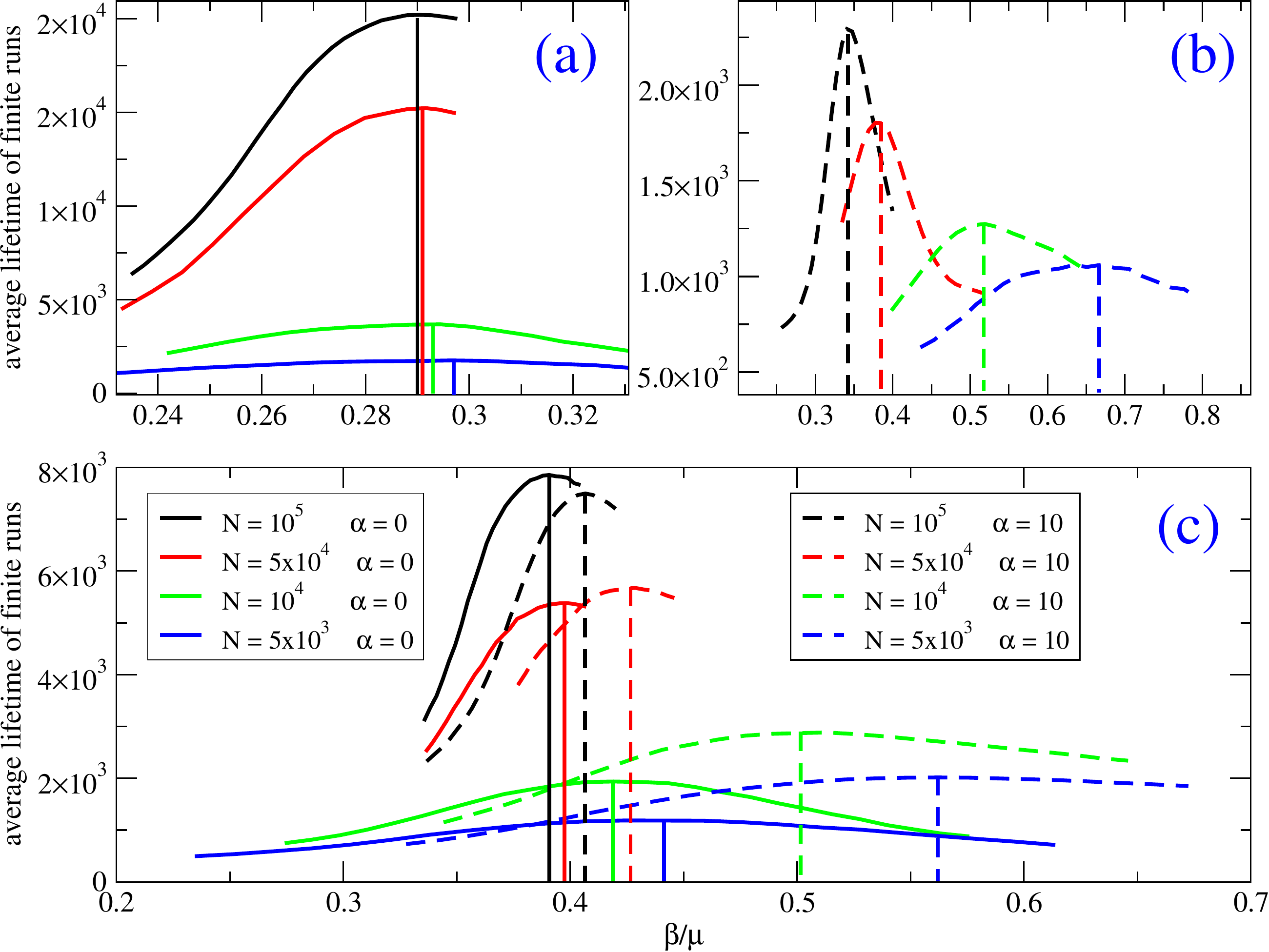}
  \caption{Analysis of finite-size effects.  We plot the average
    lifetime of non-endemic realizations of the SIS process, for
    different system sizes and 2 different values of $\alpha$. (a): ADM
    networks and $\alpha=0$. (b): ADM networks with $\alpha=10$. (c): AD
    networks. Vertical lines indicate the position of the maximum of the
    average lifetime. Model parameters: $\mu = 0.015$, $\gamma = 2$,
    $\epsilon = 10^{-3}$ and $\Delta T =\infty$. Results are averaged
    over $1000$ realizations.}
  \label{fig:finite_size_effects}
\end{figure}

In Fig.~\ref{fig:ADSISM1A1} we first explore the effect of the strength
of risk awareness, as measured by the parameter $\alpha$, in the case
$\Delta T = \infty$, i.e., when each agent is influenced by the whole
history of his/her past contacts, a situation in which awareness effects
should be maximal. We plot the steady state average fraction of
infectious nodes $\rho = \sum_a \rho_a F(a)$ as a function of
$\beta/\mu$ for three different values of $\alpha$, and evaluate the
position of the effective epidemic threshold, as measured by the peak of
the average lifetime of non-endemic realizations, see
Sec.~\ref{sec:epid-models-epid}.  Figures~\ref{fig:ADSISM1A1}c) and d)
indicate that the effect of awareness in the model ($\alpha > 0$), with
respect to the pure SIS model ($\alpha = 0$) is to reduce the fraction
$\rho$ of infectious individuals for all values of $\beta/\mu$, and
Figures~\ref{fig:ADSISM1A1}a) and b) seem to indicate in addition a
shift of the effective epidemic threshold to larger values. This effect
is more pronounced for the ADM than for the AD networks.  As this shift
of the epidemic threshold is in contradiction, at least for the AD case,
with the mean-field analysis of the previous paragraphs, we investigate
this issue in more details in Fig.~\ref{fig:finite_size_effects}, where
we show, both for the pure SIS model ($\alpha=0$) and for a positive
value of $\alpha$, the average lifetime of non-endemic realizations for
various system sizes.  Strong finite-size effects are observed,
especially for the model with awareness ($\alpha > 0$).  Fitting the
values of the effective threshold (the position of the lifetime peak)
with a law of the form
$(\beta/\mu)_N = (\beta/\mu)_\infty + A\,N^{-\nu}$, typical of
finite-size scaling analysis \cite{cardy88}, leads to a threshold in the
thermodynamic limit of $(\beta/\mu)_\infty = 0.37(3)$ for the pure SIS
model on AD networks, $(\beta/\mu)_\infty = 0.34(2)$ for AD with
$\alpha=10$ (SIS model with awareness), $(\beta/\mu)_\infty = 0.29(3)$
for ADM with $\alpha=0$ (pure SIS model) and
$(\beta/\mu)_\infty = 0.29(2)$ for ADM with $\alpha=10$. We notice here
that the extrapolations for $\alpha=0$ are less accurate and thus with
larger associated errors.  Nevertheless, with the evidence at hand, we
can conclude that, within error bars, the risk perception has no effect
on the epidemic threshold in the thermodynamic limit, in agreement with
the result from Eq.~(\ref{eq:1}), that gives a theoretical threshold
$(\beta/\mu)_c=0.366$ for the AD case. It is however noteworthy that the
effective epidemic threshold measured in finite systems can be quite
strongly affected by the awareness mechanism, even for quite large
systems, and in a particularly dramatic way for ADM networks.

\begin{figure}[t]
\centering
\includegraphics{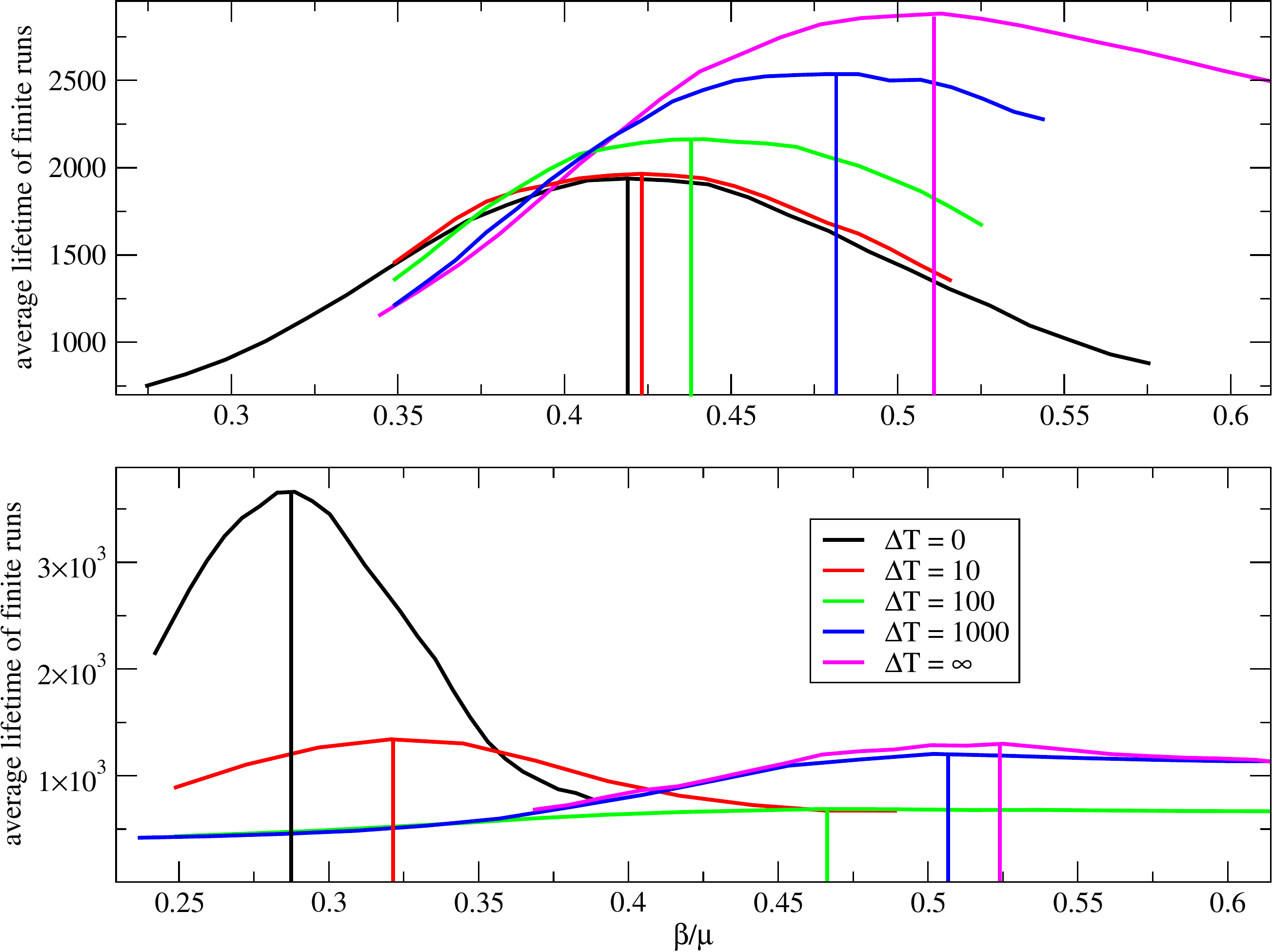} 
\caption{Effect of the local risk perception with increasing memory span
  $\Delta T$ for the SIS spreading on AD and ADM network.  (top): AD
  network.  (bottom): ADM network. Vertical lines indicate the position
  of the maximum of the average lifetime. Model parameters: $\alpha=10$,
  $\mu = 0.015$, $\gamma = 2$, $\epsilon = 10^{-3}$ and network size
  $N = 10^4$. Results are averaged over $1000$ realizations. }
\label{fig:ADSISdeltaT}
\end{figure}

We finally explore in Fig.~\ref{fig:ADSISdeltaT} the effect of a varying
memory length $\Delta T$, at fixed risk awareness strength $\alpha$. In
both AD and ADM networks, an increasing awareness temporal window shifts
the effective epidemic threshold towards larger values, up to a maximum
given by $\Delta T = \infty$, when the whole system history is
available. For the ADM networks, this effect is less clear because of
the changing height of the maximum of the lifespan when increasing
$\Delta T$. For AD networks, this result is apparently at odds with the
mean-field analysis in which $\Delta T$ is irrelevant in the stationary
state. We should notice, however, that for $\Delta T \to \infty$, the
critical point is unchanged in the thermodynamic limit with respect to
the pure SIS dynamics. Given that for $\Delta T \to \infty$ the effects
of awareness are the strongest, we expect that a finite $\Delta T$ will
not be able to change the threshold in the infinite network limit. We
can thus attribute the shifts observed to pure finite size effects.  Note that this
effect is also seen in homogeneous AD networks with uniform activity
$a$ (data not shown), observation that we can explain as follows: when
$\Delta T$ is small, the ratio of contacts with infectious
$n_I(i)_{\Delta T}$ recorded by an individual $i$ can differ
significantly from the overall ratio recorded in the whole network in
the same time window, which is equal to $\av{n_I(i)_{\Delta T}}= \rho$
(for a uniform activity). Mathematically, we have
\begin{equation}
\av{\lambda_i}= \lambda_0\, \av{\exp(-\alpha\, n_I(i)_{\Delta T})} \geq
\lambda_0 \,\exp(-\alpha\, \rho) 
\end{equation}
by concavity of the exponential function. Thus, even if locally and
temporarily some individuals perceive an overestimated prevalence of the
epidemics and reduce their probability of being infected accordingly, on
average the reduction in the transmission rate would be larger if the
ensemble average were used instead of the temporal one, and thus the
epidemics is better contained in the former case. As $\Delta T$
increases, the temporal average $n_I(i)_{\Delta T}$ becomes closer to
the ensemble one $\rho$ and the effect of awareness increases. When
$\Delta T$ is large enough compared to the time scale of variation of
the network $1/a$, the local time recording becomes equivalent to an
ensemble average, and we recover the mean-field situation.

\subsection{SIR dynamics}
\label{sec:sir-dynamics}

\subsubsection{Analytical approach}

Following an approach similar to the case of the SIS model, the SIR model
has been studied at the heterogeneous mean field level in AD networks, in
terms of a set of equations for the state of nodes with activity $a$,
which takes the form~\cite{Liu2014}
\begin{eqnarray}
  I_{a}^{t+1}&=&I_{a}^{t}- \mu \, I_{a}^{t}+ \lambda\, a\, (N_a -
  I_{a}^{t} - R_{a}^{t})\int
                 \dfrac{I_{a'}^{t}}{N}\,da' \nonumber \\
  &+& \lambda\, (N_a -
  I_{a}^{t} - R_{a}^{t} )\int
\dfrac{I_{a'}^{t}\,a'}{N}\,da'  \ ,
\label{eq:SIRAD}
\end{eqnarray}
where $N_a$ is the total number of nodes with activity $a$, and $I_a$ and
$R_a$ are the number of nodes with activity $a$ in the infectious and
recovered states, respectively. Again, a linear stability analysis
shows the presence of a threshold, which takes the same form as in the SIS case:
\begin{equation}
  \label{eq:2}
  \frac{\beta}{\mu} \geq  \frac{2 \av{a}}{\av{a} + \sqrt{\av{a^2}}}.
\end{equation}
The same expression can be obtained by a different approach, based on
the mapping of the SIR processes to bond
percolation~\cite{Starnini2014}.

Since the SIR model lacks a steady state, we cannot apply in the general
case the approach followed in the previous section.  The effects of risk
perception can be however treated theoretically for a homogeneous
network (uniform activity) in the limit $\Delta T \to \infty$, which is
defined by the effective infection probability
\begin{equation}
  \lambda(t) = \lambda_0 \,\exp\left(- \frac{\alpha}{t}\int_{0}^{t}
    \rho(\tau)\,d\tau\right) \ .   
\label{eq:10}
\end{equation}
Even this case is hard to tackle analytically, so that we consider instead a
modified model defined by the infection probability
\begin{equation}
  \lambda(t) = \lambda_0 \,\exp\left(-\alpha
    \int_{0}^{t}\rho(\tau)\,d\tau\right) \ .  
\end{equation}
In this definition the fraction of infectious seen by an individual is
no longer averaged over the memory length but rather accumulated over
the memory timespan, so that we expect stronger effects of the risk
perception with respect to Eq.~\eqref{eq:2}, if any.  The fraction of
susceptibles $s=S/N$ and the fraction of recovered $r=R/N$ in the system
obey the equations
\begin{eqnarray}
  \dfrac{ds}{dt} &=& -\lambda_0 \,\rho(t) \,s(t)\, e^{-\alpha r(t)/\mu}\\
  \dfrac{dr}{dt} &=& \mu \rho(t)
\end{eqnarray}
where in the first equation we have used the second equation to replace
$\int_{0}^{t}\rho(\tau)\,d\tau$ in $\lambda(t)$ by $(r(t) - r(0))/\mu$
(with the initial conditions $r(0)=0$).
 
Setting $\mu=1$ without loss of generality, 
the final average fraction of recovered individuals after the end of an
outbreak is given by
\begin{equation}
  r_{\infty} =   1-s(0) \exp\left(-\dfrac{\lambda_0}{\alpha}(1-e^{-\alpha
      r_{\infty}})\right) .
\end{equation}
Close to the threshold, i.e., for $r_{\infty} \sim 0$, performing an
expansion up to second order and imposing the initial condition
$\rho(0) = 1 - s(0) = 0$, we obtain the asymptotic solution
\begin{equation}
  r_{\infty} \simeq \frac{2}{ \lambda_0(\alpha+\lambda_0)} (\lambda_0 - 1),
  \label{eq:9}
\end{equation}
which leads to the critical infection rate $\lambda_0 = 1$. This means
that, as for the SIS case, the risk perception does not affect the
epidemic threshold at the mean field level, at least for a homogeneous
network. The only effect of awareness is a depression of the order
parameter $r_\infty$ with $\alpha$, as observed also in the SIS
case. The same conclusion is expected to hold for the original model of
awareness, with an infection rate of the form Eq.~(\ref{eq:10}) as in
this case the dynamics is affected to a lower extent. In analogy, for
the general case of an heterogeneous AD network, with rate infection
given by Eq.~\eqref{risk}, we expect the effects of awareness on the
epidemic threshold to be negligible at the mean-field level.

On ADM networks, the numerical analysis of the SIR model carried out in
Ref.~\cite{Sun:2015} has revealed a picture opposite to the SIS case. In
an SIR process indeed, reinfection is not possible; as a result,
repeating contacts are not useful for the diffusion of the infection.
The spread is thus favoured by the more random patterns occurring in the
memory-less (AD) case, which allows infectious nodes to contact a broader
range of different individuals and find new susceptible ones. The
epidemic threshold for SIR processes is hence higher in the ADM case
than in the AD one~\cite{Sun:2015}.

\begin{figure}[t]
  \centering \includegraphics{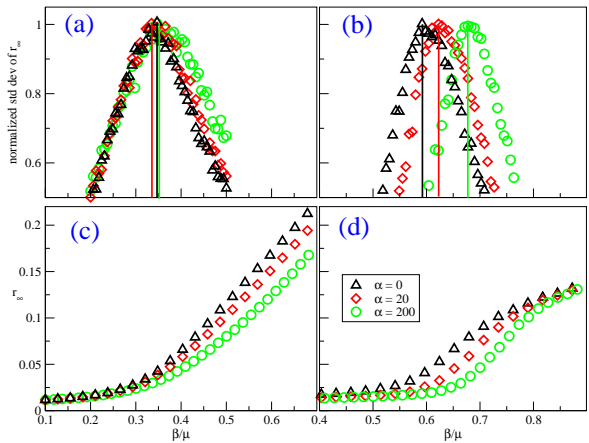}
  \caption{Effect of the local risk perception on the SIR spreading on
    AD networks and ADM networks. We plot $r_{\infty}$ and
    $\sigma_r/\sigma_r^{max}$ for different values of $\alpha$. (a):
    $\sigma_r/\sigma_r^{max}$ on AD network, (b):
    $\sigma_r/\sigma_r^{max}$ on ADM network, (c): $r_{\infty}$ on AD
    network and (d): $r_{\infty}$ on ADM network.  Vertical lines in
    subplots (a) and (b) indicate the position of the maximum of the
    order parameter variance.  Model parameters: $\Delta T=\infty$,
    $\mu = 0.015$, $\gamma = 2$, $\epsilon = 10^{-3}$ and network size
    $N = 10^5$. Results are averaged over $1000$
    realizations. }
  \label{fig:ADSIRM1A1}
\end{figure}

\subsubsection{Numerical simulations}

To study the effects of risk perception on the dynamics of a SIR
spreading process in temporal networks we resort again to numerical
simulations.  In Fig.~\ref{fig:ADSIRM1A1} we compare the effects of the
risk perception mechanism given by Eq.~\eqref{risk} for AD and ADM
networks. The spread starts with a fraction $\rho_0 = I_0/N=0.01$ of
infectious nodes chosen at random in the population and the activity
distribution is the same as in the SIS case. In the present simulations
the memory span $\Delta T$ is infinite and we compare the results
obtained for two different values of the awareness strenght $\alpha$. We
see that the effective epidemic threshold is increased for the ADM
network, whereas it seems unchanged for the AD network and around a value of 
$\beta/\mu =0.35$, an agreement with the theoretical prediction quoted in the previous section.

\begin{figure}[t]
  \centering \includegraphics{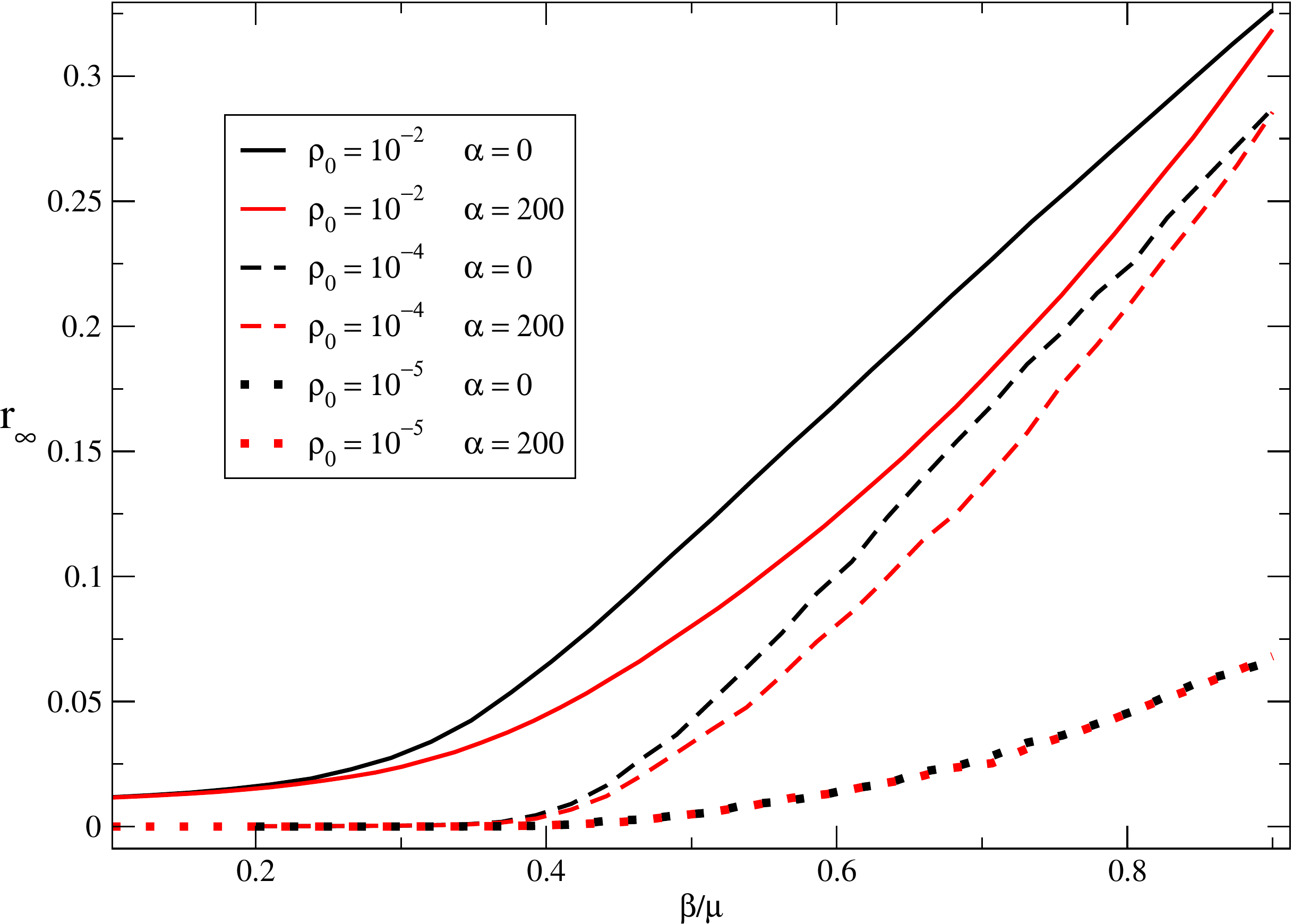}
  \caption{Effect of the initial density of infectious on the SIR model on
    AD networks for different values of the awareness strength $\alpha$
    and the initial density of infectious individuals $\rho_0$.  Model
    parameters: $\Delta T=\infty$, $\mu = 0.015$, $\gamma = 2$,
    $\epsilon = 10^{-3}$ and network size $N = 10^5$. Results are
    averaged over $1000$ realizations.}
\label{fig:ADSIRM1A1rho1}
\end{figure}

The SIR phase transition is rigorously defined for a vanishing initial
density of infectious, i.e., in the limit $\rho(0) \to 0$ and
$s(0) \to 1$, as can be seen at the mean-field level in the derivation
of Eq.~(\ref{eq:9}).  In Fig.~\ref{fig:ADSIRM1A1rho1} we explore the
effects of the initial density $\rho_0=I_0/N$ of infectious individuals
on the effect of awareness on AD networks. For large values of
$\rho_0=I_0/N$, the awareness ($\alpha > 0$) can significantly decrease
the final epidemic size, as already observed in
Fig.~\ref{fig:ADSIRM1A1}. This effect can be understood by the fact
that, for large $\rho_0$, more individuals are aware already from the
start of the spread and have therefore lower probabilities to be
infected.  At very small initial densities, on the other hand,
$r_\infty$ becomes independent of $\alpha$. This is at odds with the
result in Eq.~(\ref{eq:9}), which however was obtained within an
approximation that increases the effects of awareness. The milder form
considered in Eq.~(\ref{risk}) leads instead to an approximately
unaltered threshold, and to a prevalence independent of $\alpha$.

\begin{figure}[t]
\centering
\includegraphics{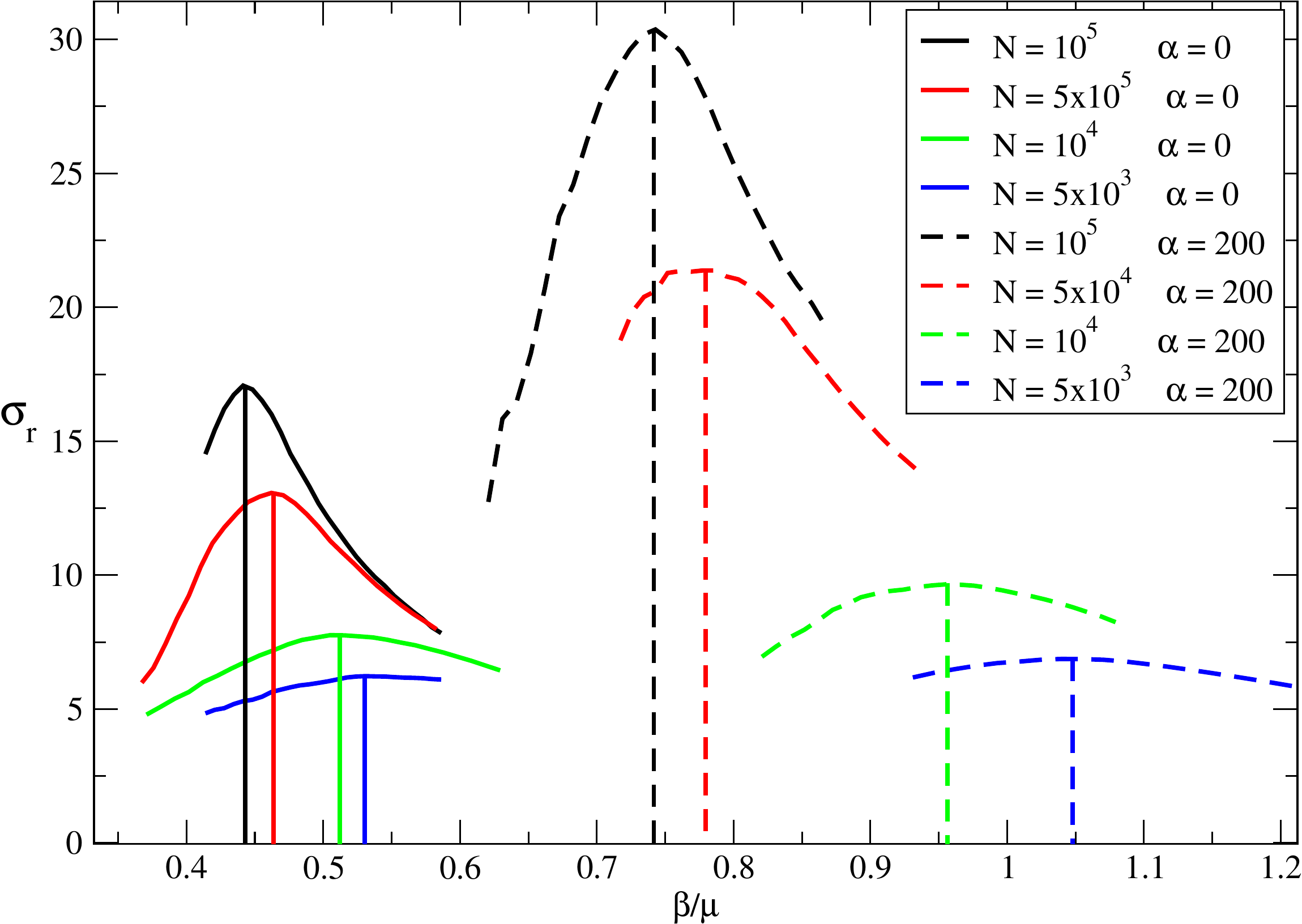} 
\caption{Finite scale effects in the SIR process on ADM. we plot
  $\sigma_r$ for different network sizes and two values of
  $\alpha$. Vertical lines indicate the position of the maximum of the
  order parameter variance. Model parameters: $\rho_0=1/N$, $\Delta T=\infty$,
  $\mu = 0.005$, $\gamma = 2$, $\epsilon = 10^{-3}$. Results are
  averaged over $10^5$ realizations.}
\label{fig:SIRfinitesize}
\end{figure}

For ADM networks, Fig.~\ref{fig:SIRfinitesize} shows the variance of the
order parameter for two different values of $\alpha$. As in the SIS
case, we see that an apparent shift of the effective epidemic threshold
is obtained, but very strong finite size effects are present even at
large size, especially for $\alpha>0$. The difference between the
effective thresholds at $\alpha >0$ and $\alpha=0$ decreases as the
system size increases, but remains quite large, making it difficult to
reach a clear conclusion on the infinite size limit.

\section{Epidemic spreading on empirical social networks}

As neither AD nor ADM networks display all the complex multi-scale
features of real contact networks, we now turn to numerical simulations
of spreading processes with and without awareness on empirical temporal
contact networks, using the datasets described in Sec.~\ref{sec:empir-soci-netw}.

\subsection{SIS dynamics}

\begin{figure}[t]
\centering
\includegraphics{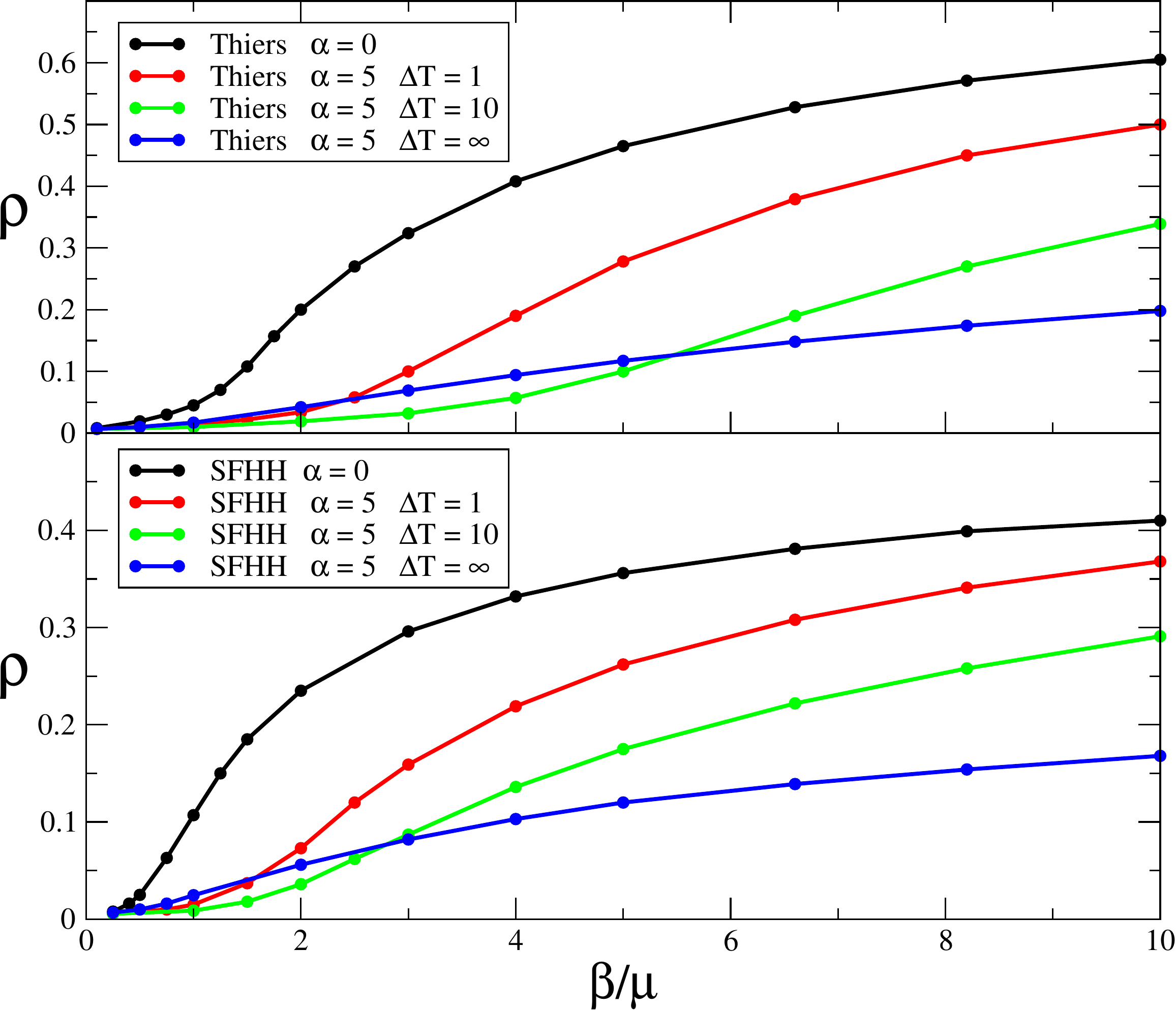}
\caption{Steady state fraction of infectious for the SIS process on both
  empirical networks, for 2 values of $\alpha$ and different values of
  $\Delta T$. Model parameters: $\mu=0.001$ for Thiers and $\mu = 0.005$
  for SFHH. Results are averaged over 1000 realizations.}
\label{fig:HSISA1}
\end{figure} 

As we saw in Sec.~\ref{sec:sis-dynamics}, the susceptibility defined to
evaluate the epidemic threshold of the SIS process is subject to strong
finite size effects. Since the empirical networks used in the present
section are quite small, we choose to focus only on the main observable
of physical interest, i.e., the average prevalence $\rho$ in the steady
state of the epidemics.

As we are interested in the influence of the structural properties of
the network, we choose to skip the nights in the datasets describing the contacts between individuals, as obviously no social
activity was recorded then, to avoid undesired extinction of the epidemic
during those periods. In order to run simulations of the SIS spreading,
we construct from the data arbitrarily long lasting periodic networks,
with the period being the recording duration (once the
nights have been removed). For both networks we define the average instantaneous degree
$\av{k} = \dfrac{1}{T_{data}}\sum_i \overline{k}_t$ where the sum runs over all the time
steps of the data, and $\overline{k}_t$ is the average degree of the
snapshot network at time $t$. We then define
$\beta/\mu = \lambda \av{k}/\mu$ as the parameter of the epidemic. For
each run, a random starting time step is chosen, and a single agent in
the same time step, if there is any, is defined as the seed of the
infection (otherwise a new starting time is chosen).

In Fig.~\ref{fig:HSISA1}, we compare the curves of the prevalence $\rho$
of the epidemics in the stationary state on both empirical networks, and
for increasing values of the memory length $\Delta T$. We can see that
an important reduction of the prevalence is occurring even for
$\Delta T=1$. This is due to the presence of many contacts of duration longer than $\Delta T$ (contrarily to the AD case): 
the awareness mechanism decreases the probability of contagion of all these contacts (and in particular of the
contacts with very long duration, which have an important role in the propagation) as soon as $\Delta T > 1$, leading to a strong
effect even in this case. At large values of the control parameter $\beta/\mu$, the effect of the awareness is stronger for increasing
values of the memory length $\Delta T$, as was observed in Sec.~\ref{sec:sis-dynamics}.
At small values of $\beta/\mu$ on the contrary, the awareness is optimum for a finite value of $\Delta T$, and the
knowledge of the whole contact history is not the best way to contain
the epidemics. While a detailed investigation of this effect lies beyond the scope of our work,
preliminary work (not shown) seem to indicate that it is linked to the use of the periodicity introduced in the data
through the repetition of the dataset.

\subsection{SIR}

\begin{figure}[t]
\centering
\includegraphics{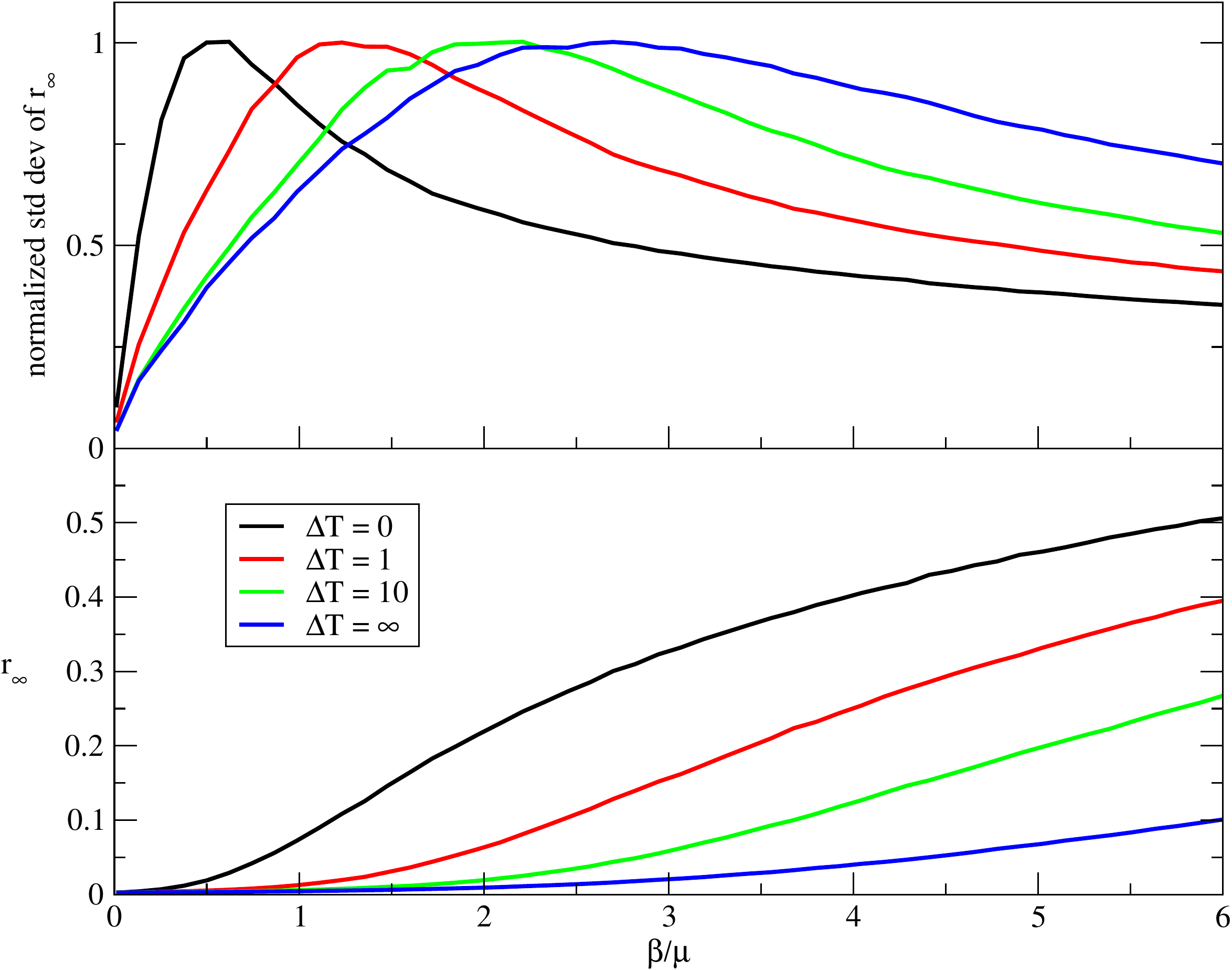}
\caption{Effect of the risk perception for different values of
  $\Delta T$ on the SIR spreading on SFHH network. (top): normalized
  standard deviation $\sigma_r/\sigma_r^{max}$. (bottom): order
  parameter $r_{\infty}$. Model parameters: $\mu =
  0.005$. $\alpha=200$. Results are averaged over $10^4$ realizations.}
\label{fig:HSIRA1}
\end{figure}

\begin{figure}[t]
\centering
\includegraphics{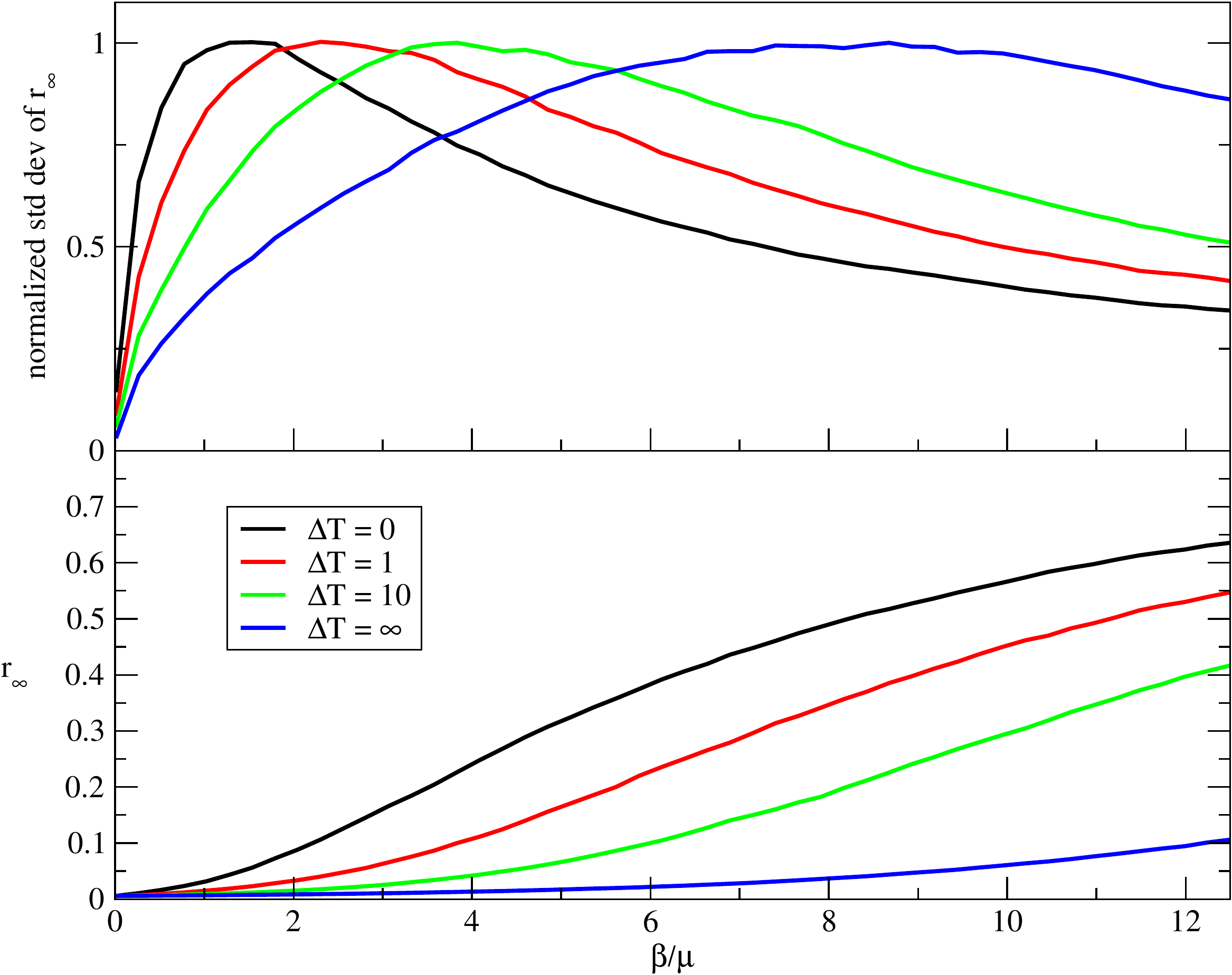}
\caption{Effect of the risk perception for different values of
  $\Delta T$ on the SIR spreading on the Thiers network.  (top):
  normalized standard deviation $\sigma_r/\sigma_r^{max}$. (bottom):
  order parameter $r_{\infty}$.  Model parameters: $\mu =
  0.001$. $\alpha=200$. Results are averaged over $10^4$ realizations.}
\label{fig:thiersSIR}
\end{figure}

In this section we study the impact of the awareness on the SIR
spreading process running on the empirical networks. In particular, we
study the effect of self protection on the fraction of recovered
individuals $r_\infty$ in the final state, and on the effective
threshold evaluated as the peak of the relative variance of $r_\infty$
defined in Eq.~\eqref{sigma}. In Fig.~\ref{fig:HSIRA1} and
\ref{fig:thiersSIR} we plot $\sigma_r$ and $r_\infty$ for different
memory length $\Delta T$, for the SFHH conference and the Thiers
highschool data respectively. We first notice that a notable effect
appears already for $\Delta T=1$, similarly to the SIS process. However,
we see that $r_\infty$ is monotonously reduced as $\Delta T$ grows and
that the effective threshold is shifted to higher values of $\beta/\mu$,
also monotonously.  It is worth noticing that the timescale of the SIR
process is much smaller than the one studied in the SIS process because
the final state is an absorbing state free of infectious agents. The
lifetime of the epidemic in this case is of the order of magnitude of
the data duration, so that the periodicity introduced by the repetition
of the dataset is not relevant anymore. Overall, we observe for both
networks an important reduction of outbreak size when people adopt a
self protecting behaviour, as well as a significant shift of the
effective epidemic threshold.

\section{Conclusion}

The implementation of immunization strategies to contain the propagation
of epidemic outbreaks in social networks is a task of paramount
importance. In this work, we have considered the effects of taking
protective measures to avoid infection in the context of social temporal
networks, a more faithful representation of the patterns of social
contacts than often considered static structures. In this context,
we have implemented a model including awareness to the propagating
disease in a temporal network, extending previous approaches defined for
static frameworks. In our model, susceptible individuals have a local
perception of the overall disease prevalence measured as the ratio of
the number of previous contacts with infectious individuals on a training
window of width $\Delta T$. An increased level of awareness induces a
reduction in the probability that a susceptible individual contracts the
disease via a contact with an infectious individual.

To explore the effects of disease awareness we have considered the
paradigmatic SIS and SIR spreading models on both synthetic temporal
networks, based in the activity driven (AD) model paradigm, and
empirical face-to-face contact networks collected by the SocioPatterns
collaboration. In the case of network models, we consider the original AD
model, and a variation, the AD model with memory (ADM), in which a
memory kernel mimics some of the non-Markovian effects observed in
real social networks.

In the case of synthetic networks, analytical and numerical results hint
that in AD networks without memory, the epidemic threshold on both SIS
and SIR models is not changed by the presence of awareness, while the
epidemic prevalence is diminished for increasing values of the parameter
$\alpha$ gauging the strength of awareness.  In the case of the ADM
model (temporal network with memory effects) on the other hand,
awareness seems to be able to shift the threshold to an increased value,
but very strong finite size effects are present: our results are
compatible with an absence of change of the epidemic threshold in the
infinite size limit, while, as for the AD case, the epidemic prevalence
is decreased.

In the case of empirical contact networks, we observe in all cases a
strong reduction of the prevalence for different values of $\alpha$ and
$\Delta T$, and an apparent shift of the effective epidemic threshold.
These empirical networks differ from the network models from two crucial
points of view. On the one hand, they have a relatively small
size. Given that important finite size effects are observed in the
models, especially in the one with memory effects, one might also expect
stronger effective shifts in such populations of limited size.  On the
other hand, AD and ADM networks lack numerous realistic features
observed in real social systems. On AD and ADM networks, contacts are
established with random nodes (even in the ADM case) so that the
perception of the density of infectious by any node is quite
homogeneous, at least in the hypothesis of a sufficiently large number
of contacts recorded (i.e., at large enough times, for
$a\Delta T \gg 1$).  This is not the case for the empirical networks,
which exhibits complex patterns such as community structures, as well as
broad distributions of contact and inter-contact durations, specific
time-scales (e.g., lunch breaks), correlated activity patterns,
etc.~\cite{Gauvin:2014}. This rich topological and temporal structure
can lead to strong heterogeneities in the local perception of the
disease. In this respect, it would be interesting to investigate the
effect of awareness in more realistic temporal network models.

Notably, the awareness mechanism, even if only local and not assuming
any global knowledge of the unfolding of the epidemics, leads to a
strong decrease of the prevalence and to shifts in the effective
epidemic threshold even at quite large size, in systems as diverse as
simple models and empirical data. Moreover, some features of empirical
contact networks, such as the broad distribution of contact durations,
seem to enhance this effect even for short-term memory awareness.
Overall, our results indicate that it would be important to take into
account awareness effects as much as possible in data-driven simulations
of epidemic spread, to study the relative role of the complex properties
of contact networks on these effects, and we hope this will stimulate
more research into this crucial topic.

\begin{acknowledgements}
  R.P.-S. acknowledgs financial support from the Spanish Government's
  MINECO, under projects FIS2013-47282-C2- 2 and FIS2016-76830-C2-1-P,
  from ICREA Academia, funded by the \textit{Generalitat de Catalunya}
  regional authorities.
\end{acknowledgements}


\begin{thebibliography}{41}%
\makeatletter
\providecommand \@ifxundefined [1]{%
 \@ifx{#1\undefined}
}%
\providecommand \@ifnum [1]{%
 \ifnum #1\expandafter \@firstoftwo
 \else \expandafter \@secondoftwo
 \fi
}%
\providecommand \@ifx [1]{%
 \ifx #1\expandafter \@firstoftwo
 \else \expandafter \@secondoftwo
 \fi
}%
\providecommand \natexlab [1]{#1}%
\providecommand \enquote  [1]{``#1''}%
\providecommand \bibnamefont  [1]{#1}%
\providecommand \bibfnamefont [1]{#1}%
\providecommand \citenamefont [1]{#1}%
\providecommand \href@noop [0]{\@secondoftwo}%
\providecommand \href [0]{\begingroup \@sanitize@url \@href}%
\providecommand \@href[1]{\@@startlink{#1}\@@href}%
\providecommand \@@href[1]{\endgroup#1\@@endlink}%
\providecommand \@sanitize@url [0]{\catcode `\\12\catcode `\$12\catcode
  `\&12\catcode `\#12\catcode `\^12\catcode `\_12\catcode `\%12\relax}%
\providecommand \@@startlink[1]{}%
\providecommand \@@endlink[0]{}%
\providecommand \url  [0]{\begingroup\@sanitize@url \@url }%
\providecommand \@url [1]{\endgroup\@href {#1}{\urlprefix }}%
\providecommand \urlprefix  [0]{URL }%
\providecommand \Eprint [0]{\href }%
\providecommand \doibase [0]{http://dx.doi.org/}%
\providecommand \selectlanguage [0]{\@gobble}%
\providecommand \bibinfo  [0]{\@secondoftwo}%
\providecommand \bibfield  [0]{\@secondoftwo}%
\providecommand \translation [1]{[#1]}%
\providecommand \BibitemOpen [0]{}%
\providecommand \bibitemStop [0]{}%
\providecommand \bibitemNoStop [0]{.\EOS\space}%
\providecommand \EOS [0]{\spacefactor3000\relax}%
\providecommand \BibitemShut  [1]{\csname bibitem#1\endcsname}%
\let\auto@bib@innerbib\@empty
\bibitem [{\citenamefont {Keeling}\ and\ \citenamefont
  {Rohani}(2008)}]{Keeling-Rohani_Book}%
  \BibitemOpen
  \bibfield  {author} {\bibinfo {author} {\bibfnamefont {M.~J.}\ \bibnamefont
  {Keeling}}\ and\ \bibinfo {author} {\bibfnamefont {P.}~\bibnamefont
  {Rohani}},\ }\href@noop {} {\emph {\bibinfo {title} {Modeling Infectious
  Diseases in Humans and Animals}}}\ (\bibinfo  {publisher} {Princeton
  University Press},\ \bibinfo {address} {Princeton},\ \bibinfo {year}
  {2008})\BibitemShut {NoStop}%
\bibitem [{\citenamefont {Anderson}\ and\ \citenamefont
  {May}(1991)}]{Anderson-May_Book}%
  \BibitemOpen
  \bibfield  {author} {\bibinfo {author} {\bibfnamefont {R.~M.}\ \bibnamefont
  {Anderson}}\ and\ \bibinfo {author} {\bibfnamefont {R.~M.}\ \bibnamefont
  {May}},\ }\href@noop {} {\emph {\bibinfo {title} {Infectious diseases of
  humans: dynamics and control}}}\ (\bibinfo  {publisher} {Oxford University
  Press},\ \bibinfo {address} {Oxford, New York},\ \bibinfo {year}
  {1991})\BibitemShut {NoStop}%
\bibitem [{\citenamefont {Newman}(2010)}]{Newman10}%
  \BibitemOpen
  \bibfield  {author} {\bibinfo {author} {\bibfnamefont {M.}~\bibnamefont
  {Newman}},\ }\href@noop {} {\emph {\bibinfo {title} {Networks: An
  Introduction}}}\ (\bibinfo  {publisher} {Oxford University Press, Inc.},\
  \bibinfo {address} {New York, NY, USA},\ \bibinfo {year} {2010})\BibitemShut
  {NoStop}%
\bibitem [{\citenamefont {Caldarelli}(2007)}]{caldarelli2007sfn}%
  \BibitemOpen
  \bibfield  {author} {\bibinfo {author} {\bibfnamefont {G.}~\bibnamefont
  {Caldarelli}},\ }\href@noop {} {\emph {\bibinfo {title} {{Scale-Free
  Networks: Complex Webs in Nature and Technology}}}}\ (\bibinfo  {publisher}
  {Oxford University Press},\ \bibinfo {address} {Oxford},\ \bibinfo {year}
  {2007})\BibitemShut {NoStop}%
\bibitem [{\citenamefont {Pastor-Satorras}\ and\ \citenamefont
  {Vespignani}(2001)}]{Pastor-Satorras:2001}%
  \BibitemOpen
  \bibfield  {author} {\bibinfo {author} {\bibfnamefont {R.}~\bibnamefont
  {Pastor-Satorras}}\ and\ \bibinfo {author} {\bibfnamefont {A.}~\bibnamefont
  {Vespignani}},\ }\href@noop {} {\bibfield  {journal} {\bibinfo  {journal}
  {Phys. Rev. Lett.}\ }\textbf {\bibinfo {volume} {86}},\ \bibinfo {pages}
  {3200} (\bibinfo {year} {2001})}\BibitemShut {NoStop}%
\bibitem [{\citenamefont {Barrat}\ \emph {et~al.}(2008)\citenamefont {Barrat},
  \citenamefont {Barthelemy},\ and\ \citenamefont {Vespignani}}]{BBV}%
  \BibitemOpen
  \bibfield  {author} {\bibinfo {author} {\bibfnamefont {A.}~\bibnamefont
  {Barrat}}, \bibinfo {author} {\bibfnamefont {M.}~\bibnamefont {Barthelemy}},
  \ and\ \bibinfo {author} {\bibfnamefont {A.}~\bibnamefont {Vespignani}},\
  }\href@noop {} {\emph {\bibinfo {title} {Dynamical processes on complex
  networks}}}\ (\bibinfo  {publisher} {Cambridge University Press
  (Cambridge)},\ \bibinfo {year} {2008})\BibitemShut {NoStop}%
\bibitem [{\citenamefont {Pastor-Satorras}\ \emph {et~al.}(2015)\citenamefont
  {Pastor-Satorras}, \citenamefont {Castellano}, \citenamefont {Mieghem},\ and\
  \citenamefont {Vespignani}}]{Pastor-Satorras:2015}%
  \BibitemOpen
  \bibfield  {author} {\bibinfo {author} {\bibfnamefont {R.}~\bibnamefont
  {Pastor-Satorras}}, \bibinfo {author} {\bibfnamefont {C.}~\bibnamefont
  {Castellano}}, \bibinfo {author} {\bibfnamefont {P.~V.}\ \bibnamefont
  {Mieghem}}, \ and\ \bibinfo {author} {\bibfnamefont {A.}~\bibnamefont
  {Vespignani}},\ }\href@noop {} {\bibfield  {journal} {\bibinfo  {journal}
  {Rev. Mod. Phys.}\ }\textbf {\bibinfo {volume} {87}},\ \bibinfo {pages} {925}
  (\bibinfo {year} {2015})}\BibitemShut {NoStop}%
\bibitem [{\citenamefont {Holme}\ and\ \citenamefont
  {Saram{\"a}ki}(2012)}]{Holme:2012}%
  \BibitemOpen
  \bibfield  {author} {\bibinfo {author} {\bibfnamefont {P.}~\bibnamefont
  {Holme}}\ and\ \bibinfo {author} {\bibfnamefont {J.}~\bibnamefont
  {Saram{\"a}ki}},\ }\href {\doibase
  http://dx.doi.org/10.1016/j.physrep.2012.03.001} {\bibfield  {journal}
  {\bibinfo  {journal} {Physics Reports}\ }\textbf {\bibinfo {volume} {519}},\
  \bibinfo {pages} {97 } (\bibinfo {year} {2012})}\BibitemShut {NoStop}%
\bibitem [{\citenamefont {Karsai}\ \emph {et~al.}(2011)\citenamefont {Karsai},
  \citenamefont {Kivel\"a}, \citenamefont {Pan}, \citenamefont {Kaski},
  \citenamefont {Kert\'esz}, \citenamefont {Barab\'asi},\ and\ \citenamefont
  {Saram\"aki}}]{Karsai:2011}%
  \BibitemOpen
  \bibfield  {author} {\bibinfo {author} {\bibfnamefont {M.}~\bibnamefont
  {Karsai}}, \bibinfo {author} {\bibfnamefont {M.}~\bibnamefont {Kivel\"a}},
  \bibinfo {author} {\bibfnamefont {R.~K.}\ \bibnamefont {Pan}}, \bibinfo
  {author} {\bibfnamefont {K.}~\bibnamefont {Kaski}}, \bibinfo {author}
  {\bibfnamefont {J.}~\bibnamefont {Kert\'esz}}, \bibinfo {author}
  {\bibfnamefont {A.-L.}\ \bibnamefont {Barab\'asi}}, \ and\ \bibinfo {author}
  {\bibfnamefont {J.}~\bibnamefont {Saram\"aki}},\ }\href {\doibase
  10.1103/PhysRevE.83.025102} {\bibfield  {journal} {\bibinfo  {journal} {Phys.
  Rev. E}\ }\textbf {\bibinfo {volume} {83}},\ \bibinfo {pages} {025102}
  (\bibinfo {year} {2011})}\BibitemShut {NoStop}%
\bibitem [{\citenamefont {Stehl{\'e}}\ \emph {et~al.}(2011)\citenamefont
  {Stehl{\'e}}, \citenamefont {Voirin}, \citenamefont {Barrat}, \citenamefont
  {Cattuto}, \citenamefont {Colizza}, \citenamefont {Isella}, \citenamefont
  {R{\'e}gis}, \citenamefont {Pinton}, \citenamefont {Khanafer}, \citenamefont
  {Van~den Broeck},\ and\ \citenamefont {Vanhems}}]{Stehle:2011}%
  \BibitemOpen
  \bibfield  {author} {\bibinfo {author} {\bibfnamefont {J.}~\bibnamefont
  {Stehl{\'e}}}, \bibinfo {author} {\bibfnamefont {N.}~\bibnamefont {Voirin}},
  \bibinfo {author} {\bibfnamefont {A.}~\bibnamefont {Barrat}}, \bibinfo
  {author} {\bibfnamefont {C.}~\bibnamefont {Cattuto}}, \bibinfo {author}
  {\bibfnamefont {V.}~\bibnamefont {Colizza}}, \bibinfo {author} {\bibfnamefont
  {L.}~\bibnamefont {Isella}}, \bibinfo {author} {\bibfnamefont
  {C.}~\bibnamefont {R{\'e}gis}}, \bibinfo {author} {\bibfnamefont {J.-F.}\
  \bibnamefont {Pinton}}, \bibinfo {author} {\bibfnamefont {N.}~\bibnamefont
  {Khanafer}}, \bibinfo {author} {\bibfnamefont {W.}~\bibnamefont {Van~den
  Broeck}}, \ and\ \bibinfo {author} {\bibfnamefont {P.}~\bibnamefont
  {Vanhems}},\ }\href {\doibase 10.1186/1741-7015-9-87} {\bibfield  {journal}
  {\bibinfo  {journal} {BMC Medicine}\ }\textbf {\bibinfo {volume} {9}},\
  \bibinfo {pages} {87} (\bibinfo {year} {2011})}\BibitemShut {NoStop}%
\bibitem [{\citenamefont {Machens}\ \emph {et~al.}(2013)\citenamefont
  {Machens}, \citenamefont {Gesualdo}, \citenamefont {Rizzo}, \citenamefont
  {Tozzi}, \citenamefont {Barrat},\ and\ \citenamefont
  {Cattuto}}]{Machens:2013}%
  \BibitemOpen
  \bibfield  {author} {\bibinfo {author} {\bibfnamefont {A.}~\bibnamefont
  {Machens}}, \bibinfo {author} {\bibfnamefont {F.}~\bibnamefont {Gesualdo}},
  \bibinfo {author} {\bibfnamefont {C.}~\bibnamefont {Rizzo}}, \bibinfo
  {author} {\bibfnamefont {A.}~\bibnamefont {Tozzi}}, \bibinfo {author}
  {\bibfnamefont {A.}~\bibnamefont {Barrat}}, \ and\ \bibinfo {author}
  {\bibfnamefont {C.}~\bibnamefont {Cattuto}},\ }\href {\doibase
  10.1186/1471-2334-13-185} {\bibfield  {journal} {\bibinfo  {journal} {BMC
  Infectious Diseases}\ }\textbf {\bibinfo {volume} {13}},\ \bibinfo {pages}
  {185} (\bibinfo {year} {2013})}\BibitemShut {NoStop}%
\bibitem [{\citenamefont {Valdano}\ \emph {et~al.}(2015)\citenamefont
  {Valdano}, \citenamefont {Ferreri}, \citenamefont {Poletto},\ and\
  \citenamefont {Colizza}}]{Valdano:2015}%
  \BibitemOpen
  \bibfield  {author} {\bibinfo {author} {\bibfnamefont {E.}~\bibnamefont
  {Valdano}}, \bibinfo {author} {\bibfnamefont {L.}~\bibnamefont {Ferreri}},
  \bibinfo {author} {\bibfnamefont {C.}~\bibnamefont {Poletto}}, \ and\
  \bibinfo {author} {\bibfnamefont {V.}~\bibnamefont {Colizza}},\ }\href@noop
  {} {\bibfield  {journal} {\bibinfo  {journal} {Phys. Rev. X}\ }\textbf
  {\bibinfo {volume} {5}},\ \bibinfo {pages} {021005} (\bibinfo {year}
  {2015})}\BibitemShut {NoStop}%
\bibitem [{\citenamefont {Holme}(2015)}]{Holme:2015}%
  \BibitemOpen
  \bibfield  {author} {\bibinfo {author} {\bibfnamefont {P.}~\bibnamefont
  {Holme}},\ }\href {\doibase 10.1140/epjb/e2015-60657-4} {\bibfield  {journal}
  {\bibinfo  {journal} {The European Physical Journal B}\ }\textbf {\bibinfo
  {volume} {88}},\ \bibinfo {pages} {234} (\bibinfo {year} {2015})}\BibitemShut
  {NoStop}%
\bibitem [{\citenamefont {Cattuto}\ \emph {et~al.}(2010)\citenamefont
  {Cattuto}, \citenamefont {Van~den Broeck}, \citenamefont {Barrat},
  \citenamefont {Colizza}, \citenamefont {Pinton},\ and\ \citenamefont
  {Vespignani}}]{Cattuto:2010}%
  \BibitemOpen
  \bibfield  {author} {\bibinfo {author} {\bibfnamefont {C.}~\bibnamefont
  {Cattuto}}, \bibinfo {author} {\bibfnamefont {W.}~\bibnamefont {Van~den
  Broeck}}, \bibinfo {author} {\bibfnamefont {A.}~\bibnamefont {Barrat}},
  \bibinfo {author} {\bibfnamefont {V.}~\bibnamefont {Colizza}}, \bibinfo
  {author} {\bibfnamefont {J.-F.}\ \bibnamefont {Pinton}}, \ and\ \bibinfo
  {author} {\bibfnamefont {A.}~\bibnamefont {Vespignani}},\ }\href {\doibase
  10.1371/journal.pone.0011596} {\bibfield  {journal} {\bibinfo  {journal}
  {PLoS ONE}\ }\textbf {\bibinfo {volume} {5}},\ \bibinfo {pages} {e11596}
  (\bibinfo {year} {2010})}\BibitemShut {NoStop}%
\bibitem [{\citenamefont {Barrat}\ \emph {et~al.}(2014)\citenamefont {Barrat},
  \citenamefont {Cattuto}, \citenamefont {Tozzi}, \citenamefont {Vanhems},\
  and\ \citenamefont {Voirin}}]{Barrat:2014}%
  \BibitemOpen
  \bibfield  {author} {\bibinfo {author} {\bibfnamefont {A.}~\bibnamefont
  {Barrat}}, \bibinfo {author} {\bibfnamefont {C.}~\bibnamefont {Cattuto}},
  \bibinfo {author} {\bibfnamefont {A.~E.}\ \bibnamefont {Tozzi}}, \bibinfo
  {author} {\bibfnamefont {P.}~\bibnamefont {Vanhems}}, \ and\ \bibinfo
  {author} {\bibfnamefont {N.}~\bibnamefont {Voirin}},\ }\href@noop {}
  {\bibfield  {journal} {\bibinfo  {journal} {Clinical Microbiology and
  Infection}\ }\textbf {\bibinfo {volume} {20}},\ \bibinfo {pages} {10}
  (\bibinfo {year} {2014})}\BibitemShut {NoStop}%
\bibitem [{\citenamefont {Barab\`asi}(2005)}]{Barabasi:2005}%
  \BibitemOpen
  \bibfield  {author} {\bibinfo {author} {\bibfnamefont {A.-L.}\ \bibnamefont
  {Barab\`asi}},\ }\href@noop {} {\bibfield  {journal} {\bibinfo  {journal}
  {Nature}\ }\textbf {\bibinfo {volume} {435}},\ \bibinfo {pages} {207}
  (\bibinfo {year} {2005})}\BibitemShut {NoStop}%
\bibitem [{\citenamefont {Pastor-Satorras}\ and\ \citenamefont
  {Vespignani}(2002)}]{Pastor:2002}%
  \BibitemOpen
  \bibfield  {author} {\bibinfo {author} {\bibfnamefont {R.}~\bibnamefont
  {Pastor-Satorras}}\ and\ \bibinfo {author} {\bibfnamefont {A.}~\bibnamefont
  {Vespignani}},\ }\href@noop {} {\bibfield  {journal} {\bibinfo  {journal}
  {Phys. Rev. E}\ }\textbf {\bibinfo {volume} {65}},\ \bibinfo {pages} {036104}
  (\bibinfo {year} {2002})}\BibitemShut {NoStop}%
\bibitem [{\citenamefont {Cohen}\ \emph {et~al.}(2003)\citenamefont {Cohen},
  \citenamefont {Havlin},\ and\ \citenamefont {ben Avraham}}]{Cohen:2003}%
  \BibitemOpen
  \bibfield  {author} {\bibinfo {author} {\bibfnamefont {R.}~\bibnamefont
  {Cohen}}, \bibinfo {author} {\bibfnamefont {S.}~\bibnamefont {Havlin}}, \
  and\ \bibinfo {author} {\bibfnamefont {D.}~\bibnamefont {ben Avraham}},\
  }\href@noop {} {\bibfield  {journal} {\bibinfo  {journal} {Phys. Rev. Lett.}\
  }\textbf {\bibinfo {volume} {91}},\ \bibinfo {pages} {247901} (\bibinfo
  {year} {2003})}\BibitemShut {NoStop}%
\bibitem [{\citenamefont {Funk}\ \emph {et~al.}(2010)\citenamefont {Funk},
  \citenamefont {Salath\'e},\ and\ \citenamefont {Jansen}}]{Funk:2010}%
  \BibitemOpen
  \bibfield  {author} {\bibinfo {author} {\bibfnamefont {S.}~\bibnamefont
  {Funk}}, \bibinfo {author} {\bibfnamefont {M.}~\bibnamefont {Salath\'e}}, \
  and\ \bibinfo {author} {\bibfnamefont {V.}~\bibnamefont {Jansen}},\
  }\href@noop {} {\bibfield  {journal} {\bibinfo  {journal} {J R Soc
  Interface}\ }\textbf {\bibinfo {volume} {7}},\ \bibinfo {pages} {1247}
  (\bibinfo {year} {2010})}\BibitemShut {NoStop}%
\bibitem [{\citenamefont {Gross}\ and\ \citenamefont
  {Blasius}(2008)}]{Gross:2008}%
  \BibitemOpen
  \bibfield  {author} {\bibinfo {author} {\bibfnamefont {T.}~\bibnamefont
  {Gross}}\ and\ \bibinfo {author} {\bibfnamefont {B.}~\bibnamefont
  {Blasius}},\ }\href {\doibase 10.1098/rsif.2007.1229} {\bibfield  {journal}
  {\bibinfo  {journal} {Journal of The Royal Society Interface}\ }\textbf
  {\bibinfo {volume} {5}},\ \bibinfo {pages} {259} (\bibinfo {year}
  {2008})}\BibitemShut {NoStop}%
\bibitem [{\citenamefont {Perra}\ \emph {et~al.}(2011)\citenamefont {Perra},
  \citenamefont {Balcan}, \citenamefont {Gon\c{c}alves},\ and\ \citenamefont
  {Vespignani}}]{Perra:2011}%
  \BibitemOpen
  \bibfield  {author} {\bibinfo {author} {\bibfnamefont {N.}~\bibnamefont
  {Perra}}, \bibinfo {author} {\bibfnamefont {D.}~\bibnamefont {Balcan}},
  \bibinfo {author} {\bibfnamefont {B.}~\bibnamefont {Gon\c{c}alves}}, \ and\
  \bibinfo {author} {\bibfnamefont {A.}~\bibnamefont {Vespignani}},\
  }\href@noop {} {\bibfield  {journal} {\bibinfo  {journal} {PLoS ONE}\
  }\textbf {\bibinfo {volume} {6}},\ \bibinfo {pages} {e23084} (\bibinfo {year}
  {2011})}\BibitemShut {NoStop}%
\bibitem [{\citenamefont {Granell}\ \emph {et~al.}(2013)\citenamefont
  {Granell}, \citenamefont {G\'{o}mez},\ and\ \citenamefont
  {Arenas}}]{Granell:2013}%
  \BibitemOpen
  \bibfield  {author} {\bibinfo {author} {\bibfnamefont {C.}~\bibnamefont
  {Granell}}, \bibinfo {author} {\bibfnamefont {S.}~\bibnamefont {G\'{o}mez}},
  \ and\ \bibinfo {author} {\bibfnamefont {A.}~\bibnamefont {Arenas}},\ }\href
  {\doibase 10.1103/PhysRevLett.111.128701} {\bibfield  {journal} {\bibinfo
  {journal} {Physical Review Letters}\ }\textbf {\bibinfo {volume} {111}},\
  \bibinfo {pages} {128701} (\bibinfo {year} {2013})}\BibitemShut {NoStop}%
\bibitem [{\citenamefont {Massaro}\ and\ \citenamefont
  {Bagnoli}(2014)}]{Massaro:2014}%
  \BibitemOpen
  \bibfield  {author} {\bibinfo {author} {\bibfnamefont {E.}~\bibnamefont
  {Massaro}}\ and\ \bibinfo {author} {\bibfnamefont {F.}~\bibnamefont
  {Bagnoli}},\ }\href {\doibase 10.1103/PhysRevE.90.052817} {\bibfield
  {journal} {\bibinfo  {journal} {Phys. Rev. E}\ }\textbf {\bibinfo {volume}
  {90}},\ \bibinfo {pages} {052817} (\bibinfo {year} {2014})}\BibitemShut
  {NoStop}%
\bibitem [{\citenamefont {Bagnoli}\ \emph {et~al.}(2007)\citenamefont
  {Bagnoli}, \citenamefont {Lio},\ and\ \citenamefont
  {Sguanci}}]{Bagnoli:2007}%
  \BibitemOpen
  \bibfield  {author} {\bibinfo {author} {\bibfnamefont {F.}~\bibnamefont
  {Bagnoli}}, \bibinfo {author} {\bibfnamefont {P.}~\bibnamefont {Lio}}, \ and\
  \bibinfo {author} {\bibfnamefont {L.}~\bibnamefont {Sguanci}},\ }\href@noop
  {} {\bibfield  {journal} {\bibinfo  {journal} {Phys Rev E}\ }\textbf
  {\bibinfo {volume} {76}},\ \bibinfo {pages} {061904} (\bibinfo {year}
  {2007})}\BibitemShut {NoStop}%
\bibitem [{\citenamefont {Kotnis}\ and\ \citenamefont
  {Kuri}(2013)}]{Kotnis:2013}%
  \BibitemOpen
  \bibfield  {author} {\bibinfo {author} {\bibfnamefont {B.}~\bibnamefont
  {Kotnis}}\ and\ \bibinfo {author} {\bibfnamefont {J.}~\bibnamefont {Kuri}},\
  }\href {\doibase 10.1103/PhysRevE.87.062810} {\bibfield  {journal} {\bibinfo
  {journal} {Phys. Rev. E}\ }\textbf {\bibinfo {volume} {87}},\ \bibinfo
  {pages} {062810} (\bibinfo {year} {2013})}\BibitemShut {NoStop}%
\bibitem [{\citenamefont {Rizzo}\ \emph {et~al.}(2014)\citenamefont {Rizzo},
  \citenamefont {Frasca},\ and\ \citenamefont {Porfiri}}]{Rizzo:2014}%
  \BibitemOpen
  \bibfield  {author} {\bibinfo {author} {\bibfnamefont {A.}~\bibnamefont
  {Rizzo}}, \bibinfo {author} {\bibfnamefont {M.}~\bibnamefont {Frasca}}, \
  and\ \bibinfo {author} {\bibfnamefont {M.}~\bibnamefont {Porfiri}},\ }\href
  {\doibase 10.1103/PhysRevE.90.042801} {\bibfield  {journal} {\bibinfo
  {journal} {Phys. Rev. E}\ }\textbf {\bibinfo {volume} {90}},\ \bibinfo
  {pages} {042801} (\bibinfo {year} {2014})}\BibitemShut {NoStop}%
\bibitem [{\citenamefont {{Cao, Lang}}(2014)}]{Cao:2014}%
  \BibitemOpen
  \bibfield  {author} {\bibinfo {author} {\bibnamefont {{Cao, Lang}}},\ }\href
  {\doibase 10.1140/epjb/e2014-50422-8} {\bibfield  {journal} {\bibinfo
  {journal} {Eur. Phys. J. B}\ }\textbf {\bibinfo {volume} {87}},\ \bibinfo
  {pages} {225} (\bibinfo {year} {2014})}\BibitemShut {NoStop}%
\bibitem [{\citenamefont {Perra}\ \emph {et~al.}(2012)\citenamefont {Perra},
  \citenamefont {Gon\c{c}alves}, \citenamefont {Pastor-Satorras},\ and\
  \citenamefont {Vespignani}}]{Perra:2012}%
  \BibitemOpen
  \bibfield  {author} {\bibinfo {author} {\bibfnamefont {N.}~\bibnamefont
  {Perra}}, \bibinfo {author} {\bibfnamefont {B.}~\bibnamefont
  {Gon\c{c}alves}}, \bibinfo {author} {\bibfnamefont {R.}~\bibnamefont
  {Pastor-Satorras}}, \ and\ \bibinfo {author} {\bibfnamefont {A.}~\bibnamefont
  {Vespignani}},\ }\href@noop {} {\bibfield  {journal} {\bibinfo  {journal}
  {Scientific reports}\ }\textbf {\bibinfo {volume} {2}},\ \bibinfo {pages}
  {469} (\bibinfo {year} {2012})}\BibitemShut {NoStop}%
\bibitem [{\citenamefont {Starnini}\ and\ \citenamefont
  {Pastor-Satorras}(2013)}]{PhysRevE.87.062807}%
  \BibitemOpen
  \bibfield  {author} {\bibinfo {author} {\bibfnamefont {M.}~\bibnamefont
  {Starnini}}\ and\ \bibinfo {author} {\bibfnamefont {R.}~\bibnamefont
  {Pastor-Satorras}},\ }\href {\doibase 10.1103/PhysRevE.87.062807} {\bibfield
  {journal} {\bibinfo  {journal} {Phys. Rev. E}\ }\textbf {\bibinfo {volume}
  {87}},\ \bibinfo {pages} {062807} (\bibinfo {year} {2013})}\BibitemShut
  {NoStop}%
\bibitem [{\citenamefont {Karsai}\ \emph {et~al.}(2014)\citenamefont {Karsai},
  \citenamefont {Perra},\ and\ \citenamefont {Vespignani}}]{Karsai:2014}%
  \BibitemOpen
  \bibfield  {author} {\bibinfo {author} {\bibfnamefont {M.}~\bibnamefont
  {Karsai}}, \bibinfo {author} {\bibfnamefont {N.}~\bibnamefont {Perra}}, \
  and\ \bibinfo {author} {\bibfnamefont {A.}~\bibnamefont {Vespignani}},\
  }\href {\doibase 10.1038/srep04001} {\bibfield  {journal} {\bibinfo
  {journal} {Sci Rep}\ }\textbf {\bibinfo {volume} {4}},\ \bibinfo {pages}
  {4001} (\bibinfo {year} {2014})}\BibitemShut {NoStop}%
\bibitem [{Soc()}]{SocioPatterns}%
  \BibitemOpen
  \href@noop {} {\enquote {\bibinfo {title} {Sociopatterns collaboration},}\
  }\bibinfo {howpublished} {\url{www.sociopatterns.org}},\ \bibinfo {note}
  {accessed 13 Feb 2017}\BibitemShut {NoStop}%
\bibitem [{Sch()}]{SchoolDataset}%
  \BibitemOpen
  \href@noop {} {\enquote {\bibinfo {title} {Sociopatterns dataset: High school
  dynamic contact networks},}\ }\bibinfo {howpublished}
  {\url{www.sociopatterns.org/datasets/high-school-dynamic-contact-networks}},\
  \bibinfo {note} {accessed 13 Feb 2017}\BibitemShut {NoStop}%
\bibitem [{\citenamefont {Fournet}\ and\ \citenamefont
  {Barrat}(2014)}]{Fournet:2014}%
  \BibitemOpen
  \bibfield  {author} {\bibinfo {author} {\bibfnamefont {J.}~\bibnamefont
  {Fournet}}\ and\ \bibinfo {author} {\bibfnamefont {A.}~\bibnamefont
  {Barrat}},\ }\href {\doibase 10.1371/journal.pone.0107878} {\bibfield
  {journal} {\bibinfo  {journal} {PLOS ONE}\ }\textbf {\bibinfo {volume} {9}},\
  \bibinfo {pages} {1} (\bibinfo {year} {2014})}\BibitemShut {NoStop}%
\bibitem [{\citenamefont {Bogu{\~n}{\'a}}\ \emph {et~al.}(2013)\citenamefont
  {Bogu{\~n}{\'a}}, \citenamefont {Castellano},\ and\ \citenamefont
  {Pastor-Satorras}}]{Boguna:2013}%
  \BibitemOpen
  \bibfield  {author} {\bibinfo {author} {\bibfnamefont {M.}~\bibnamefont
  {Bogu{\~n}{\'a}}}, \bibinfo {author} {\bibfnamefont {C.}~\bibnamefont
  {Castellano}}, \ and\ \bibinfo {author} {\bibfnamefont {R.}~\bibnamefont
  {Pastor-Satorras}},\ }\href@noop {} {\bibfield  {journal} {\bibinfo
  {journal} {Phys. Rev. Lett.}\ }\textbf {\bibinfo {volume} {111}},\ \bibinfo
  {pages} {068701} (\bibinfo {year} {2013})}\BibitemShut {NoStop}%
\bibitem [{\citenamefont {Mata}\ \emph {et~al.}(2015)\citenamefont {Mata},
  \citenamefont {Bogu\~n\'a}, \citenamefont {Castellano},\ and\ \citenamefont
  {Pastor-Satorras}}]{Mata15}%
  \BibitemOpen
  \bibfield  {author} {\bibinfo {author} {\bibfnamefont {A.~S.}\ \bibnamefont
  {Mata}}, \bibinfo {author} {\bibfnamefont {M.}~\bibnamefont {Bogu\~n\'a}},
  \bibinfo {author} {\bibfnamefont {C.}~\bibnamefont {Castellano}}, \ and\
  \bibinfo {author} {\bibfnamefont {R.}~\bibnamefont {Pastor-Satorras}},\
  }\href {\doibase 10.1103/PhysRevE.91.052117} {\bibfield  {journal} {\bibinfo
  {journal} {Phys. Rev. E}\ }\textbf {\bibinfo {volume} {91}},\ \bibinfo
  {pages} {052117} (\bibinfo {year} {2015})}\BibitemShut {NoStop}%
\bibitem [{\citenamefont {Castellano}\ and\ \citenamefont
  {Pastor-Satorras}(2016)}]{Castellano2016}%
  \BibitemOpen
  \bibfield  {author} {\bibinfo {author} {\bibfnamefont {C.}~\bibnamefont
  {Castellano}}\ and\ \bibinfo {author} {\bibfnamefont {R.}~\bibnamefont
  {Pastor-Satorras}},\ }\href {\doibase 10.1140/epjb/e2016-60953-5} {\bibfield
  {journal} {\bibinfo  {journal} {Eur. Phys. J. B}\ }\textbf {\bibinfo {volume}
  {89}},\ \bibinfo {pages} {243} (\bibinfo {year} {2016})}\BibitemShut
  {NoStop}%
\bibitem [{\citenamefont {Sun}\ \emph {et~al.}(2015)\citenamefont {Sun},
  \citenamefont {Baronchelli},\ and\ \citenamefont {Perra}}]{Sun:2015}%
  \BibitemOpen
  \bibfield  {author} {\bibinfo {author} {\bibfnamefont {K.}~\bibnamefont
  {Sun}}, \bibinfo {author} {\bibfnamefont {A.}~\bibnamefont {Baronchelli}}, \
  and\ \bibinfo {author} {\bibfnamefont {N.}~\bibnamefont {Perra}},\
  }\href@noop {} {\bibfield  {journal} {\bibinfo  {journal} {Eur. Phys. J. B}\
  }\textbf {\bibinfo {volume} {88}},\ \bibinfo {pages} {326} (\bibinfo {year}
  {2015})}\BibitemShut {NoStop}%
\bibitem [{\citenamefont {Cardy}(1988)}]{cardy88}%
  \BibitemOpen
  \bibinfo {editor} {\bibfnamefont {J.~L.}\ \bibnamefont {Cardy}},\ ed.,\
  \href@noop {} {\emph {\bibinfo {title} {Finite Size Scaling}}},\ \bibinfo
  {series} {Current Physics-Sources and Comments}, Vol.~\bibinfo {volume} {2}\
  (\bibinfo  {publisher} {North Holland},\ \bibinfo {address} {Amsterdam},\
  \bibinfo {year} {1988})\BibitemShut {NoStop}%
\bibitem [{\citenamefont {Liu}\ \emph {et~al.}(2014)\citenamefont {Liu},
  \citenamefont {Perra}, \citenamefont {Karsai},\ and\ \citenamefont
  {Vespignani}}]{Liu2014}%
  \BibitemOpen
  \bibfield  {author} {\bibinfo {author} {\bibfnamefont {S.}~\bibnamefont
  {Liu}}, \bibinfo {author} {\bibfnamefont {N.}~\bibnamefont {Perra}}, \bibinfo
  {author} {\bibfnamefont {M.}~\bibnamefont {Karsai}}, \ and\ \bibinfo {author}
  {\bibfnamefont {A.}~\bibnamefont {Vespignani}},\ }\href {\doibase
  10.1103/PhysRevLett.112.118702} {\bibfield  {journal} {\bibinfo  {journal}
  {Physical Review Letters}\ }\textbf {\bibinfo {volume} {112}},\ \bibinfo
  {pages} {118702} (\bibinfo {year} {2014})}\BibitemShut {NoStop}%
\bibitem [{\citenamefont {Starnini}\ and\ \citenamefont
  {Pastor-Satorras}(2014)}]{Starnini2014}%
  \BibitemOpen
  \bibfield  {author} {\bibinfo {author} {\bibfnamefont {M.}~\bibnamefont
  {Starnini}}\ and\ \bibinfo {author} {\bibfnamefont {R.}~\bibnamefont
  {Pastor-Satorras}},\ }\href {\doibase 10.1103/PhysRevE.89.032807} {\bibfield
  {journal} {\bibinfo  {journal} {Physical Review E}\ }\textbf {\bibinfo
  {volume} {89}},\ \bibinfo {pages} {032807} (\bibinfo {year}
  {2014})}\BibitemShut {NoStop}%
\bibitem [{\citenamefont {Gauvin}\ \emph {et~al.}(2014)\citenamefont {Gauvin},
  \citenamefont {Panisson},\ and\ \citenamefont {Cattuto}}]{Gauvin:2014}%
  \BibitemOpen
  \bibfield  {author} {\bibinfo {author} {\bibfnamefont {L.}~\bibnamefont
  {Gauvin}}, \bibinfo {author} {\bibfnamefont {A.}~\bibnamefont {Panisson}}, \
  and\ \bibinfo {author} {\bibfnamefont {C.}~\bibnamefont {Cattuto}},\
  }\href@noop {} {\bibfield  {journal} {\bibinfo  {journal} {PLoS ONE}\
  }\textbf {\bibinfo {volume} {9}} (\bibinfo {year} {2014})}\BibitemShut
  {NoStop}%
\end{thebibliography}
\end{document}